\documentclass[aps,prfluids,preprint,showpacs,longbibliography, notitlepage,superscriptaddress]{revtex4-2} 
\usepackage{amsmath,amssymb,amsfonts}%
\usepackage{amsthm}%
\usepackage{mathrsfs}%
\usepackage{mathtools}
\usepackage{floatrow}
\usepackage{graphicx}
\usepackage{color, soul}
\usepackage{mdframed}
\usepackage{braket}
\usepackage{svg}
\usepackage{wasysym}
\usepackage[colorlinks=true,citecolor=blue,linkcolor=black]{hyperref}

\newcommand{\ts}{\textsubscript}

\newcommand{\p}{\phi}

\begin{document}

\title[Article Title]{Controlling Droplets at the Tips of Fibers}
%











\author{Mengfei He}
\email{mhe100@syr.edu}
\thanks{These authors contributed equally to this work.}
\affiliation{Department of Physics, Syracuse University, Syracuse, NY 13244}
\affiliation{BioInspired Syracuse: Institute for Material and Living Systems, Syracuse University, Syracuse, NY 13244}

\author{Samay Hulikal}
\email{shulikal@syr.edu}
\thanks{These authors contributed equally to this work.}
\affiliation{Department of Physics, Syracuse University, Syracuse, NY 13244}
\affiliation{BioInspired Syracuse: Institute for Material and Living Systems, Syracuse University, Syracuse, NY 13244}

\author{Marianna L. Marquardt}
\affiliation{Department of Physics, Syracuse University, Syracuse, NY 13244}
\affiliation{BioInspired Syracuse: Institute for Material and Living Systems, Syracuse University, Syracuse, NY 13244}
\affiliation{Department of Physics, St.~Olaf College, Northfield, MN 55057}

\author{Hao Jiang}
\affiliation{BioInspired Syracuse: Institute for Material and Living Systems, Syracuse University, Syracuse, NY 13244}
\affiliation{Department of Mechanical and Aerospace Engineering, Syracuse University, Syracuse, NY 13244}

\author{Anupam Pandey}
\affiliation{BioInspired Syracuse: Institute for Material and Living Systems, Syracuse University, Syracuse, NY 13244}
\affiliation{Department of Mechanical and Aerospace Engineering, Syracuse University, Syracuse, NY 13244}

\author{Teng Zhang}
\affiliation{BioInspired Syracuse: Institute for Material and Living Systems, Syracuse University, Syracuse, NY 13244}
\affiliation{Department of Mechanical and Aerospace Engineering, Syracuse University, Syracuse, NY 13244}

\author{Christian D. Santangelo}
\affiliation{Department of Physics, Syracuse University, Syracuse, NY 13244}
\affiliation{BioInspired Syracuse: Institute for Material and Living Systems, Syracuse University, Syracuse, NY 13244}

\author{Joseph D. Paulsen}
\affiliation{Department of Physics, Syracuse University, Syracuse, NY 13244}
\affiliation{BioInspired Syracuse: Institute for Material and Living Systems, Syracuse University, Syracuse, NY 13244}

\begin{abstract}
Many 
complex wetting behaviors of fibrous materials 
are rooted in the behaviors of individual droplets attached to pairs of fibers. 
Here, we study the splitting of a droplet held between the tips of two cylindrical fibers.  
We discover a sharp transition between two post-rupture states, navigated by 
changing the angle between the rods, in agreement with our bifurcation analysis. 
Depinning of the bridge contact line can lead to 
a much larger asymmetry between the volume of liquid left on each rod. 
This second scenario enables 
the near-complete transfer of an aqueous glycerol droplet 
between two identical vinylpolysiloxane fibers. 
We leverage this response in a device that uses a ruck to pass a droplet along a train of fibers, a proof-of-concept for the geometric control of droplets on deformable, architected surfaces. 
\end{abstract}

\maketitle


\section{Introduction}
Many materials in nature and industry are composed from collections of woven or aggregated fibers, or are decorated with hairs on their surface. 
When liquid invades the interstitial space between the fibers, diverse responses can result from a competition between elasticity and wetting, including clumping, matting, and self-assembly~\cite{Bico18,Duprat22}. 
Yet, in a dense array of hairs, the channels between the hairs can prevent the liquid interface from penetrating into the structure~\cite{cassie_baxter1944}. This is particularly relevant in many biological settings, such as 
the hairy pads of insects~\cite{eisner2000, schoreder2018}, snails~\cite{kochova2014}, geckos~\cite{huber2005} and frogs~\cite{endlein2014}, allowing them to adhere to or walk on wet or mucus-covered surfaces~\cite{hu2010}. There, the interaction of a liquid interface with the \emph{tips} of the micro-hairs is prevailing, if not more crucial, 
which raises the question: how does a liquid that sits atop a hairy surface behave as the material underneath is curled, twisted, or stretched? 

Before confronting multifiber, multidroplet systems, one may start by unraveling the
fate of a single liquid bridge as it is pulled apart by the tips of two circular rods. 
The study of liquid bridges with fixed contact lines dates back to Plateau~\cite{plateau1873} and has received continuous attention throughout the centuries~\cite{bostwick2015,meseguer2001, meseguer2003,li2016}. 
Surprisingly, a simple variation on this setup, in which one tilts one of the bridge ends and then pulls them apart, has been neglected. 
We show that the outcome of the liquid bridge rupture may be dramatically altered by small variations in the angle of the rods (Fig.~\ref{fig:angle}a), in agreement with a bifurcation analysis. 

Building on these results, we demonstrate how the system geometry and a depinning contact line can be coupled to enable a moderately non-wetting polymer (receding contact angle $\sim60^{\circ}$) to achieve a near-complete transfer ($99.995\%$) of a Newtonian liquid droplet from one fiber to another, a feat that is typically only achieved by functionalizing either the liquid~\cite{chen2021} or the solid surface~\cite{qian2011,chen2013,chen2016,jamali2021}. 
We harnesses this \emph{geometric control} to pass a droplet down a train of fibers, highlighting the potential for soft metamaterials to control the transport of liquids in novel ways.

\section{Results}

\subsection{Experiment}
\begin{figure}[tb]
	\includegraphics[width=0.8\textwidth]{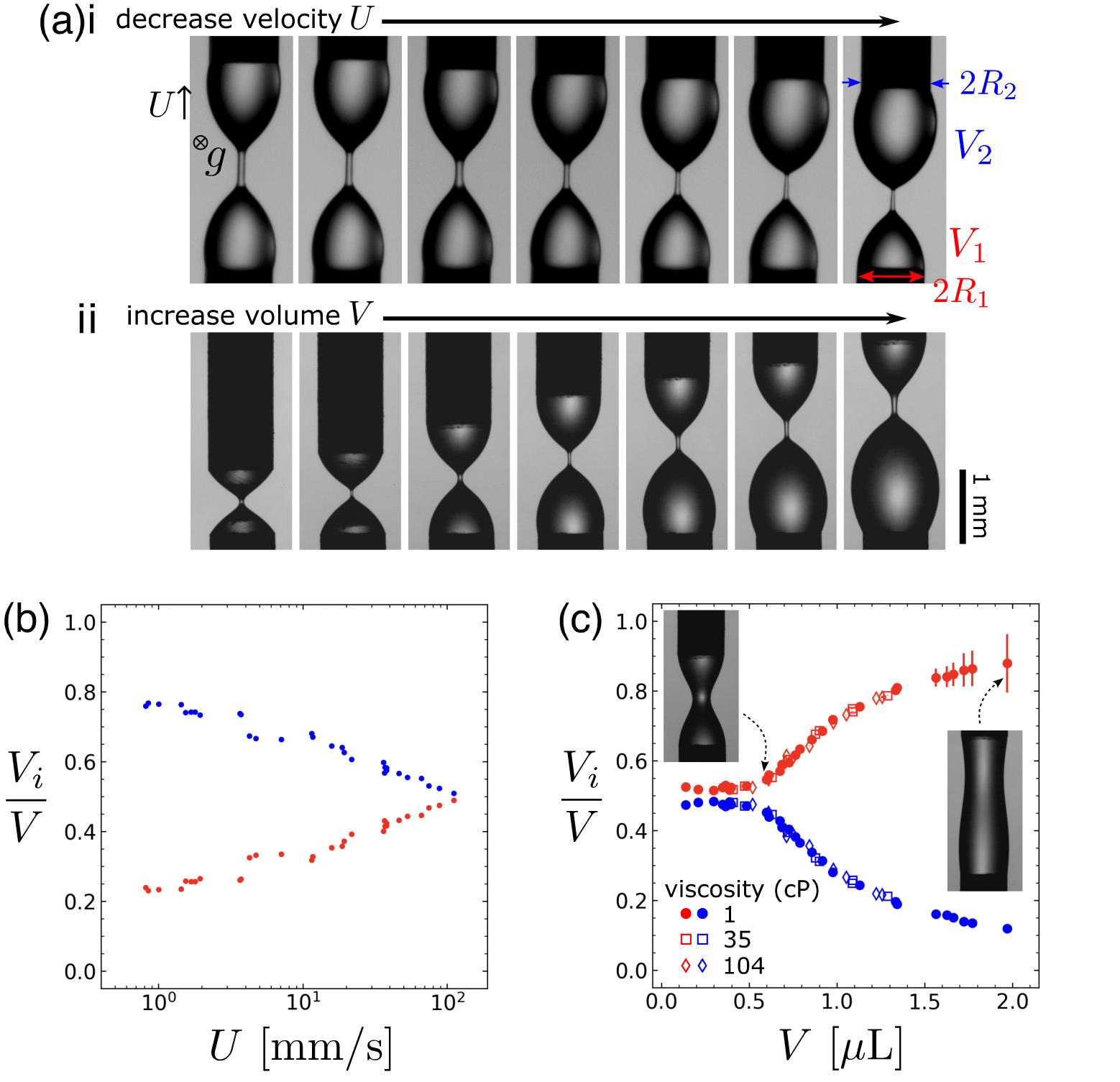}
	\caption{
	Symmetry breaking in splitting a liquid bridge between two identical rods.
	(a) Snapshots of a liquid bridge pinching off from two separating rods into droplets of volumes $V_1$ and $V_2$, for a range of i) separation velocities $U$ and ii) total volumes $V$. Symmetry is gradually broken when $U$ decreases or $V$ increases.
	(b) Fractional volumes of the separate droplets in the post-rupture state, versus $U$. 
	(c) Fractional volumes of the separate droplets in the post-rupture state, versus $U$.
	Inset: corresponding last stable states. 
	}
	\label{fig:volume}
\end{figure}

In the experiment, we form a liquid bridge of de-ionized water (surface tension $\gamma=72$ mN/m, density $\rho = 1000$ kg/m$^3$) between the circular faces of two nearly identical solid pins (Vermont Gage A series, with radii $R_1=460$ $\mu$m, $R_2=470$ $\mu$m). 
We then draw the pins apart at speed $U$. 
When the gap reaches a threshold size, the liquid bridge destabilizes and the liquid separates into two parts. 
We record this pinch-off process using a high-speed camera (Phantom Miro Lab 340). 
Our experiments are performed at a small Bond number $Bo = \rho g R_1^2/\gamma \sim 10^{-2}$, where the droplets are negligibly deformed by gravity. 
To further ensure the minimal influence of gravity, we orient our setup horizontally (gravity points into the paper in Fig.~\ref{fig:volume}a).

Figure~\ref{fig:volume}a(i) shows repeated experiments at decreasing separation velocities, $U$, with snapshots taken approximately $0.1$ ms before the liquid bridge breaks.  
At a high separation speed of $U\sim 100$ mm/s, the liquid bridge narrows at its midpoint and pinches off into two nearly symmetric droplets.  
This is intuitive, given the setup's symmetry of the boundary conditions.
However, the symmetry is gradually broken when the process is slowed down. 
At a low separation speed ($U\sim 0.1$ mm/s, last image in the sequence), there is a significant difference in the sizes of the top and bottom droplets after they break apart. 
To quantify this effect, we measure the volume, $V_1$, of the droplet remaining on the bottom rod, and that on the top rod, $V_2$, versus $U$.  Figure~\ref{fig:volume}b shows that the volume left on each fiber varies approximately logarithmically with the velocity, $U$. 

A similar broken symmetry is observed by increasing the total volume of the liquid. The image sequence of Fig.~\ref{fig:volume}a(ii) shows the uneven growth of the top and bottom droplet when the total volume of the liquid bridge increases. 
Figure~\ref{fig:volume}c also shows that starting from an even split, the volume distribution is first insensitive to the added volume when the total volume $V \lesssim 0.5$ $\mu$L. Then, for $V \gtrsim 0.5$ $\mu$L, the portion of the liquid left on each fiber becomes more and more disparate upon further increases in the initial volume, $V$. 

To investigate a possible role of the viscosity, we vary the droplet viscosity over two orders of magnitude by using mixtures of glycerol and water. 
Surprisingly, even though liquid pinch-off is a dynamic process, the amount of liquid left on each rod in the final state is observed to be independent of the droplet viscosity (different symbols in Fig.~\ref{fig:volume}c). 
We emphasize that the fluid velocities generated near the neck during the pinch-off are many orders of magnitude faster than the relatively slow velocity of the needles, $U$. 

Using high-speed imaging, we are able to trace the profile of the dynamic liquid bridge back to its marginally-stable state, where it is on the verge of destabilizing (which we call ``last stable'' hereafter). By examining the corresponding last stable states for each experiment, we find that the transition at $V \approx 0.5$ $\mu$L corresponds to the point where the liquid interface meets the face of the rod at an approximately $90^{\circ}$ angle at the last stable state (inset, Fig.~\ref{fig:volume}c).
Below this volume, the liquid bridge splits nearly evenly between the rods. 

\begin{figure}[tb]
	\includegraphics[width=0.8\textwidth]{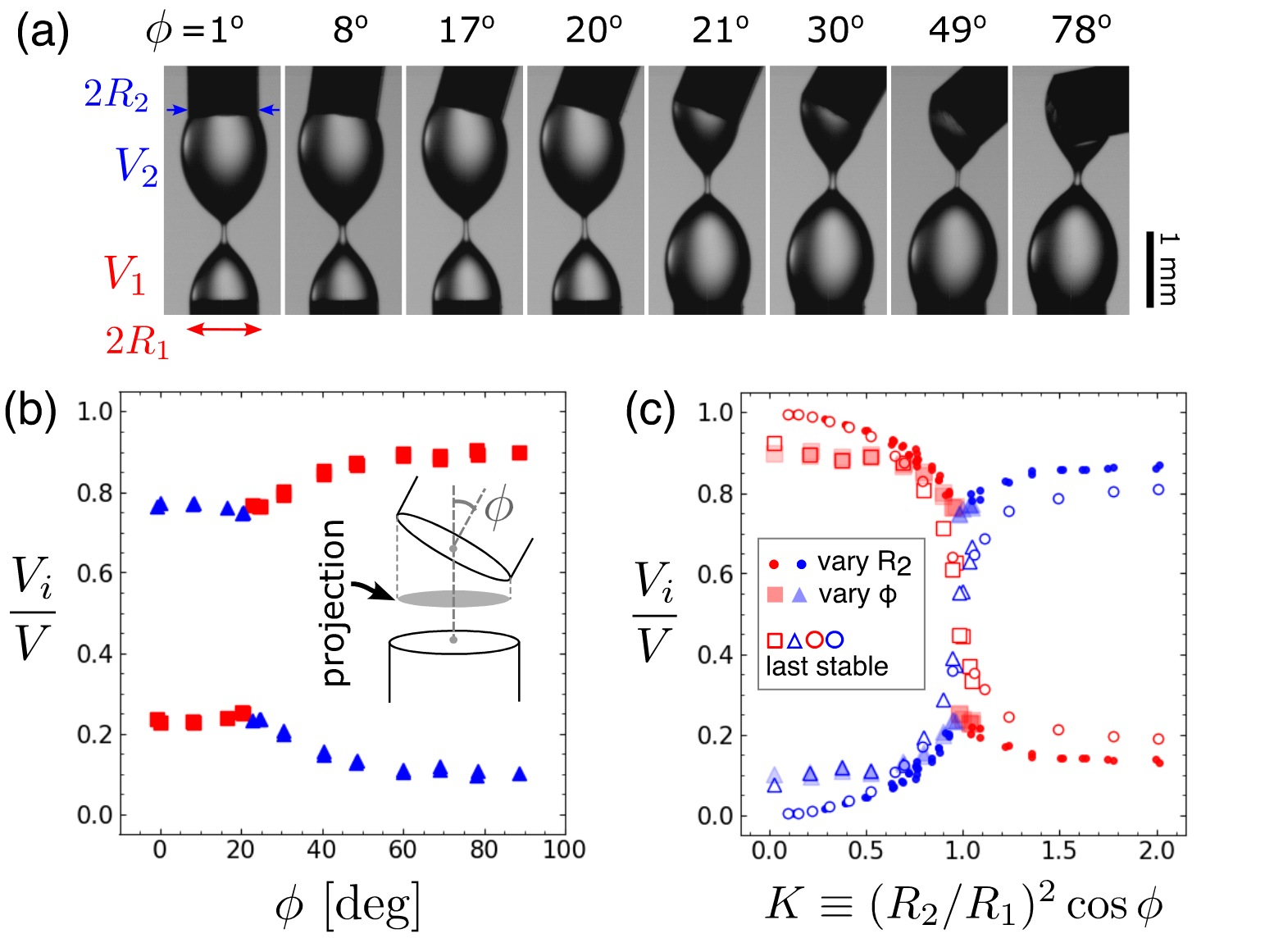}
	\caption{
    Splitting a droplet between two angled rods.
	(a) Snapshots of a liquid bridge pinching off for a range of inclination angles $\phi$. A sharp transition occurs near $\phi \approx 20^{\circ}$.  
    (b) Fractional volumes of the separate droplets in the post-rupture state, versus $\phi$. 
    (c) Fractional volumes versus the projected area ratio $K$ for the same data shown in (b) (transparent), as well as for a pair of rods at $\phi=0$ with varying $R_2$ (dots).  Open markers: last stable states for both $\phi=0$ (circles) and $\phi>0$ (squares/triangles).
	}
	\label{fig:angle}
\end{figure}

The above results identify a strong asymmetry in both the last stable state and the final state, for the \emph{quasistatic} splitting of a liquid bridge with large volume. What determines the uneven distribution? 
To investigate this question, Fig.~\ref{fig:angle}a shows a series of experiments at sequentially larger relative angles $\phi$, using the same set of rods and keeping the total volume $V$ constant.
Since the rotated rod has a slightly larger radius, $R_2>R_1$, 
the top rod initially collects a larger fraction of the liquid, consistent with what was observed at $\phi=0^{\circ}$ in Fig.~\ref{fig:volume}. 
Strikingly, as $\phi$ increases to surpass a threshold of $\phi\ts{thresh}\approx20^{\circ}$, the system goes through a sudden inversion so that the bottom rod retains more liquid.  
Figure~\ref{fig:angle}b shows that there is a significant jump in the volume distribution ($V_1$ versus $V_2$) at an abrupt, discrete transition. 
At yet larger angles, the contact line partially moves to the side wall (last panel, Fig.~\ref{fig:angle}a). Despite this complication, the bottom rod still collects a larger fraction of liquid.

\subsection{Projected area hypothesis} 
To rationalize this behavior, we start by proposing an intuitive picture of this transition.
At $\phi = 0$, it is reasonable to expect that, regardless of the details of the dynamic pinch-off process, the larger rod will end up with a larger droplet~\cite{slobozhanin1995}. For $\phi > 0$, the end of a static bridge should be narrower toward where it joins onto the tilted rod. One may thus expect that the \textit{projected} area of the rod -- which shrinks with increasing $\phi$ -- could play an important role (see schematic in Fig.~\ref{fig:angle}b).
This motivates us to consider the ratio $K$ of the projected contact areas as the key control parameter: 
\begin{equation}
K\equiv(R_2/R_1)^2\cos\phi.
\label{eq:K}
\end{equation}

At $K=1$, the projected area of the larger, tilted rod becomes equal to that of the smaller rod, which suggests a transition at $\cos^{-1} (R_1^2/R_2^2)$. As a first test of this intuition, plugging in values for $R_1$, $R_2$ for the case shown in Fig.~\ref{fig:angle}a gives a transition at $18.9^{\circ}$, close to the measured transition at $20^{\circ}$.

This way of thinking rests on the following suggested equivalence: changing the projected area of the rod (by tilting it) has the same effect as changing the actual area of an \textit{nontilted} rod. To test this idea, we perform further experiments where we split a drop using aligned rods ($\phi=0$) while systematically varying the radius of the top rod, $R_2$. We plot the final volume proportions $V_i/V$ as a function of the area ratio for the aligned rods in Fig.~\ref{fig:angle}c (dots). 
In the same plot, we again present the data for rods at angles from Fig.~\ref{fig:angle}b, now using the projected area ratio $K$ as the new variable (transparent squares/triangles). 
The transitions for both protocols are found to occur at $K = 1$. 
This result supports our proposed picture wherein the projected area of the tilted rod should be compared with the area of the untilted rod, to determine which rod receives a larger droplet in the final state. 

\begin{figure}[tb]
	\includegraphics[width=0.8\textwidth]{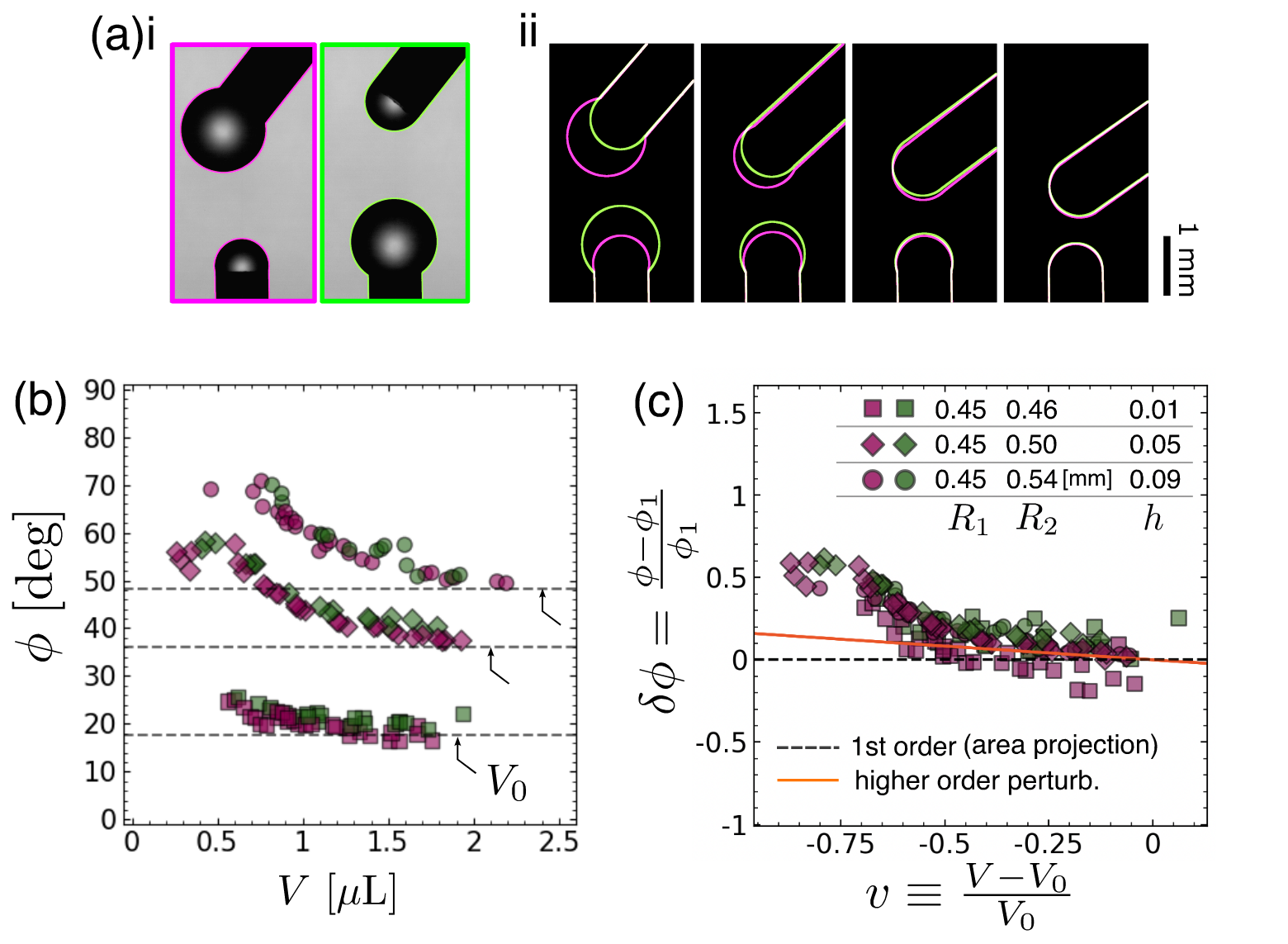}
	\caption{
    Threshold angles for different droplet volumes and rod sizes. 
	(a) i: Two possible outcomes near the threshold angle. 
	ii: Superimposed outlines of the bifurcated states at threshold angles, as the total volume $V$ is gradually reduced. 
	(b) Measured $\phi$ near the bifurcations, for three sets of rods size ratios. Green: the bottom rod receives a larger droplet.  Pink: the bottom rod receives a smaller droplet.  Horizontal dashed lines: perturbation theory Eq.~\ref{eq:firstorder}.  Arrows: corresponding Plateau limit $V=V_0$ where the perturbation theory is asymptotically accurate. 
	(c) Data collapse in $\delta \phi$-$v$ space.
    Solid orange line: higher order perturbation theory.
	}
	\label{fig:crit}
\end{figure}

\subsection{Threshold angle} 
We now test our projected area hypothesis (Eq.~\ref{eq:K}) across a wider range of rod sizes and droplet volumes. We perform a series of splitting and coalescing cycles while adjusting $\phi$ each time to hone in on the precise transition point, $\phi\ts{thresh}$. The hallmark of sitting at $\phi\ts{thresh}$ is the ability to obtain two distinct end-states with an imperceptible change in initial conditions. Figure~\ref{fig:crit}a(i) shows an example of such a fine-tuned state; we can then extract the profiles and superimpose them on one another (purple versus green outlines) to highlight the two possible states when the relative rod angle is in close proximity to $\phi\ts{thresh}$. By allowing the water to slowly evaporate, we can map out $\phi\ts{thresh}$ as a function of the liquid volume $V$, which we measure in-situ from the images of the split droplets. As the volume decreases, it takes a larger inclination for the transition to occur, while the volume disparity in the bi-modal end states decreases, as can be seen in the increasing overlap of the outlines (Fig.~\ref{fig:crit}a(ii)). We collect these data in Fig.~\ref{fig:crit}b, showing that $\phi\ts{thresh}$ decreases with increasing $V$ for three values of $R_2/R_1$ that we tested.

\subsection{Role of dynamics} 
The intricate process of a liquid bridge dynamically pinching into separate droplets occurs at multiple scales~\cite{eggers2015book}. One may wonder: does this dynamics significantly redistribute the liquid across the bridge neck, or is the splitting fraction largely set already at the static shape determined by the size and relative angle of the two rods? 
To associate a well-defined fraction of volume to each rod at the last stable state, we slice the liquid bridge with an imaginary plane at the neck of the bridge (see SI). 
We show additional data in Fig.~\ref{fig:angle}c where the volumes are measured from the last stable bridge, $V_i/V$ (open circles for $\phi=0$; open squares/triangles for $\phi>0$). These data show a close agreement with the droplets eventually separated from the liquid bridge. 
These results show that only near $K=1$ do the pinch-off dynamics play a significant role in redistributing the liquid. 

\subsection{Bifurcation theory} 

\begin{figure}[tb]
	\includegraphics[width=0.9 \textwidth]{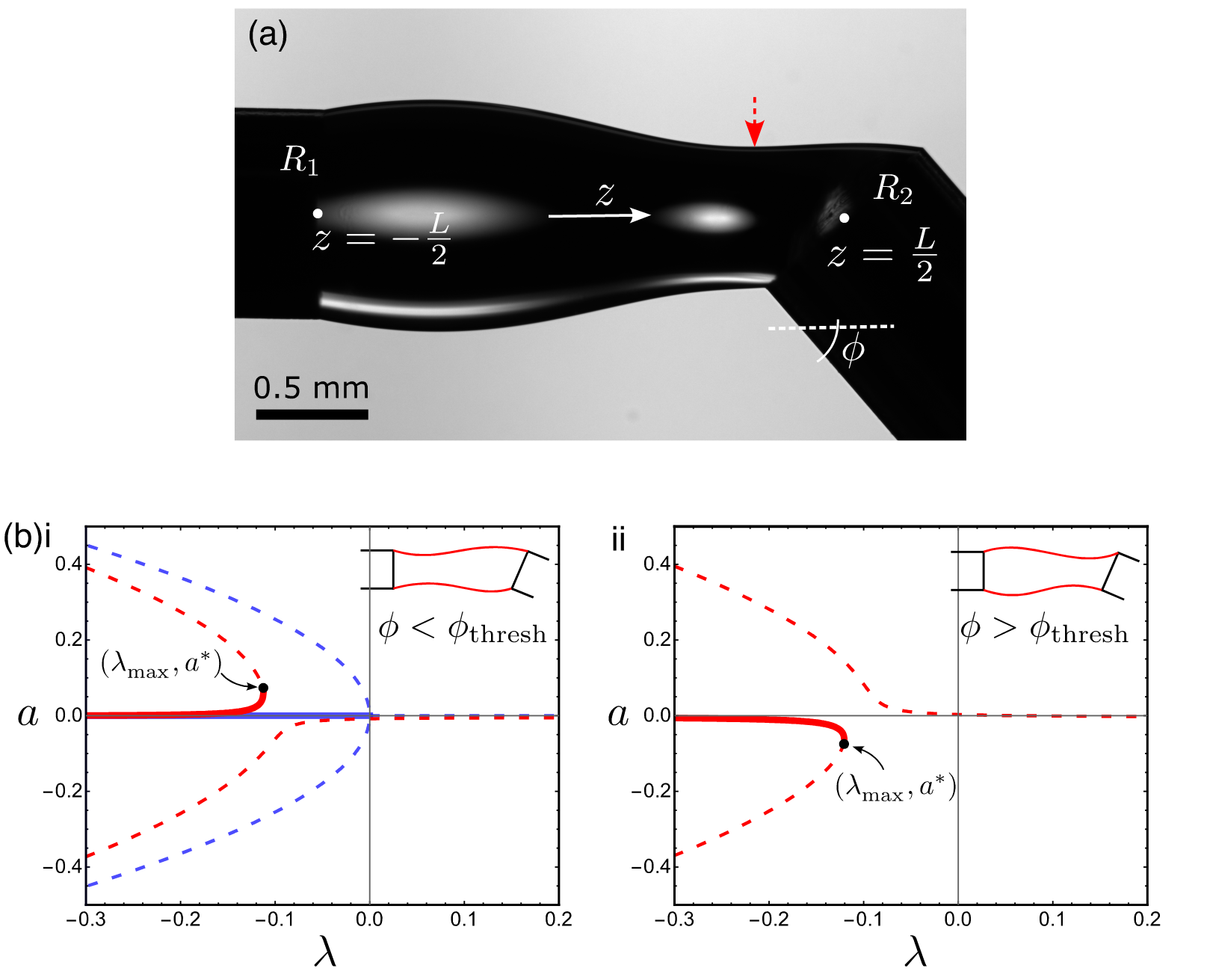}
	\caption{
    Bifurcation analysis of tilted bridges. (a) Photograph of a liquid bridge spanning two rod faces with radii $R_{1}$ and $R_{2}$, separated by $L$ and tilted by $\phi$. 
    Red arrow: position of the bridge neck. 
    (b) The amplitude, $a$, of the antisymmetric deformation as a function of the length parameter, $\lambda$, for $\phi=0$ (blue) and $\phi>0$ (red).
    The tilt angle $\phi$ controls the two unfoldings (i and ii, red) of a perfect subcritical pitchfork bifurcation (i, blue) with the change occurring across $\phi_{\text{thresh}}$. 
The location of the neck changes drastically for the two unfoldings (inset schematic).
Stable (unstable) branches are depicted as solid (dashed) curves, with a switch occurring at turning points $(\lambda_{\text{max}},a^{*})$ (black circles). 
The angled bridge bifurcations are plotted with $h=0.053$ and $v=-0.1$.
	}
	\label{fig:pitchforkbif}
\end{figure}
We have seen so far that the outcome of the droplet-splitting process is captured reasonably well by the last stable shape before the onset of the dynamic pinch-off instability. 
Motivated by this observation, we obtain an approximation of the threshold angle $\phi_{\text{thresh}}$ from the last stable shape of the liquid bridge. 

Under cylindrical coordinates $(z, \theta)$, force equilibrium at low Bond numbers requires that the profile $F(z,\theta)$ of a static liquid bridge satisfy:
\begin{align}
\mathcal{M}[F(z,\theta)]=\text{const.},
\label{eq:const_curvature}
\end{align}
where $\mathcal{M}/2$ is the mean curvature of the profile~\cite{bostwick2015}. 
Since the difference between the radii of the two rods is small, we solve Eq.~\ref{eq:const_curvature} perturbatively around a reference cylinder of radius $R_{0}=(R_{1}+R_{2}) /2$ and length $L$, with its two face centers located at $z=\pm L/2$ (Fig.~\ref{fig:pitchforkbif}a).   
At the cusp of collapse, the Plateau limit, where the length-to-diameter ratio $L/(2R_0)$ equals $\pi$, the reference cylinder has a volume of $V_{0} = 2\pi^{2} R_{0}^{3}$. 
To lowest order, the cylinder loses stability along a mode $\sin(2 \pi z /L)$ that is antisymmetric about the midplane $z=0$~\cite{atreya2002}. 
This mode characterizes the dominant symmetry-breaking deformation of the reference cylinder towards a shape $F(z,\theta)=R_{0}(1+a \sin(2 \pi z /L))$ upon varying the angle of tilt $\phi$, volume $V$, and rod radii away from their values in the reference cylinder. 
We define dimensionless parameters: 
\begin{align}
\lambda &\equiv L /(2R_{0})-\pi, \nonumber\\
h &\equiv(R_{2}-R_{1}) /(R_{2}+R_{1}), \nonumber\\
v &\equiv(V-V_{0}) /V_{0},\nonumber
\end{align}
to represent these deviations in the length, rod radii, and volume of the reference cylinder, respectively. 


Bifurcation theory enables us to express the amplitude $a$ of the antisymmetric deformation as a series expansion in the parameters $\lambda$, $h$, $v$, and $\phi$. As detailed in the SI, we present an analytical scheme utilizing the Lyapunov-Schmidt decomposition~\cite{textbook,lowgrav} to solve Eq.~\ref{eq:const_curvature}~\cite{meseguer2001, meseguer2003,atreya2002,chen1992}, while tilting the circular top support with all contact lines pinned to the edges of the supports.
The procedure yields a bifurcation equation as a power series in $(\lambda,h,\phi,v;a)$:
\begin{align}
G(\lambda,h,\phi,v;a)\equiv&\frac{2 h}{\sqrt{2+2 \pi ^2}}\nonumber\\
-&\frac{\phi ^2}{4 \sqrt{2+2 \pi ^2}} +\frac{3 a\lambda -\pi  av}{\sqrt{2+2 \pi ^2}}\nonumber\\
+&\frac{3 \pi  a^3}{2 \sqrt{2+2 \pi ^2}}-\frac{15 \pi ^2 a^2 h}{8 \sqrt{2} \left(1+\pi ^2\right)^{3/2}}\nonumber\\
+&\dots=0.
\label{eq:bifurcation}
\end{align}
Equation \ref{eq:bifurcation} has been truncated for brevity and the full equation used in our analysis can be found in the SI. 

Plotting the solutions of Eq.~\ref{eq:bifurcation} for $a$ as a function of $\lambda$ elucidates the mechanism for the sharp inversion of the volume distribution upon an angle change, presented in Fig.~\ref{fig:angle}.  First, for a cylindrical bridge, setting ($h=\phi=v=0$) produces a perfect subcritical pitchfork bifurcation~\cite{atreya2002} (blue, Fig.~\ref{fig:pitchforkbif}b(i)), and the stability is lost at the Plateau limit $\lambda=0$. 
Non-zero values of $h,\phi,v$, however, lead to two possible unfoldings of the perfect pitchfork bifurcation (red, Fig.~\ref{fig:pitchforkbif}b), depending on if the tilt of the top support surpasses a threshold $\phi_\ts{thresh}$.
As the two rods are pulled apart, the stability of the liquid bridge is lost at turning points $(\lambda_\ts{max},a^*)$ (black points, Fig.~\ref{fig:pitchforkbif}b). 
More significantly, the amplitude $a^*$ of the bridge profile switches its sign at $\phi=\phi_\ts{thresh}$, giving rise to a \emph{finite jump} of the neck position (red arrow, Fig.~\ref{fig:pitchforkbif}a) of the liquid bridge and therefore to a drastic change of the partitioned volumes.

To obtain the threshold angle $\phi_\ts{thresh}$, we effect a perturbation expansion:
\begin{align}
\phi_{\text{thresh}}&=\phi_{1}\epsilon+\phi_{2}\epsilon^{1+p_{2}}+\dots \nonumber\\
\lambda_\ts{max}&=\lambda_{1}\epsilon+\lambda_{2}\epsilon^{1+q_{2}}+\dots,\nonumber
\end{align}
for positive integers of $p_2$ and $q_2$, where the small quantity $\epsilon\sim10^{-1}$ tracks the strength of the deviations set by our experiments: $v\sim \epsilon$ and $h\sim\epsilon^2$.
At the turning points, $\partial a /\partial\lambda\to \infty$.  
We enforce the equivalent condition, $\partial G /\partial a=0$, so that the bridge is at the cusp of collapse. 
The Newton-Puiseux algorithm \cite{pui} provides a systematic procedure for obtaining the roots of the bifurcation equation (Eq. \ref{eq:bifurcation}) in powers of $\epsilon$ and, to leading order, we obtain $a^{*}=\epsilon^{2/3} \left( \frac{1}{12\pi} \right) ^{1/3}(8h-\phi_{1}^{2})^{1/3}$ (see SI).
Therefore, up to order $\epsilon^{2/3}$ the amplitude $a^*$ switches its sign at $\phi_{1}=\sqrt{8h}$, so that to the first order of $\epsilon$,
\begin{equation}\label{eq:firstorder}
\phi_\ts{thresh}=\sqrt{8h}.
\end{equation}

When $R_{2} /R_{1}\approx 1$, it follows from Eq. \ref{eq:firstorder} that $(R_{2} /R_{1})^{2}\cos \phi_{1} \approx 1-4\left( R_{2} /R_{1}-1 \right)^{2} /3+\mathcal{O}((R_{2} /R_{1}-1)^{3})$ so that $K\approx 1$ from Eq. \ref{eq:K}, validating our hypothesis that the area projection is the leading cause for the volume distribution transition. 
In Fig.~\ref{fig:crit}b, we plot Eq.~\ref{eq:firstorder} (dashed lines), and show that despite neglecting higher order terms and the dynamics of pinch-off, our bifurcation analysis successfully predicts different transition angles for different pairs of rods. As expected, the predictions from the bifurcation analysis become asymptotically accurate for volumes approaching the Plateau limits of $V_0 \approx 1.9$ $\mu$L, $V_0 \approx 2.1$ $\mu$L, and $V_0 \approx 2.4$ $\mu$L, with $R_{2}$ increasing in this order (arrows, Fig.~\ref{fig:crit}b).  
However, the theoretical threshold angle is volume-independent at this order which deviates from our data at small volumes.

To capture the volume dependence of the threshold angle, we continue the perturbation expansion to higher orders.  
The next order correction is given by $p_2=1$ and $\phi_2 =-\sqrt{2h}v/3$, so that to order $\epsilon^2$,
\begin{equation}\label{eq:secondorder}
\phi_\ts{thresh} =\sqrt{8h}-\frac{\sqrt{2h}}{3}v.
\end{equation}
Taking into account the correction term with volume $V=1.23$ $\mu$L , we obtain $\phi_{\text{thresh}}=20.06^{\circ}$, in excellent agreement with the measured transition (Fig.~\ref{fig:angle}a).  
Further, Eqs.~\ref{eq:firstorder} and~\ref{eq:secondorder} suggest that the deviation of the threshold angle from its lowest-order value, $\delta \phi \equiv (\phi_{\text{thresh}}-\phi_1)/\phi_1$, is proportional to the volume deviation $v$.  
We show in Fig.~\ref{fig:crit}c that in the $\delta \phi$-$v$ space, our data indeed collapse, demarcating a narrow boundary between the two distributions (pink versus green).  In the same space we have also plotted the first order (Eq.~\ref{eq:firstorder}) and the second order (Eq.~\ref{eq:secondorder}) solutions as the dashed and the solid lines, respectively.
Remarkably, a perturbation expansion around a straight cylinder gives a quantitative explanation for the observed transition boundary of a liquid bridge down to a volume of $40\%$ of a Plateau cylinder.  
For even smaller volumes $v\rightarrow-1^{+}$, higher order corrections play a non-negligible role, and the current expansion is insufficient. 
A detailed account of our bifurcation analysis and perturbation procedure can be found in the SI.

\begin{figure}[tb]
	\centering
	\includegraphics[width=0.82\textwidth]{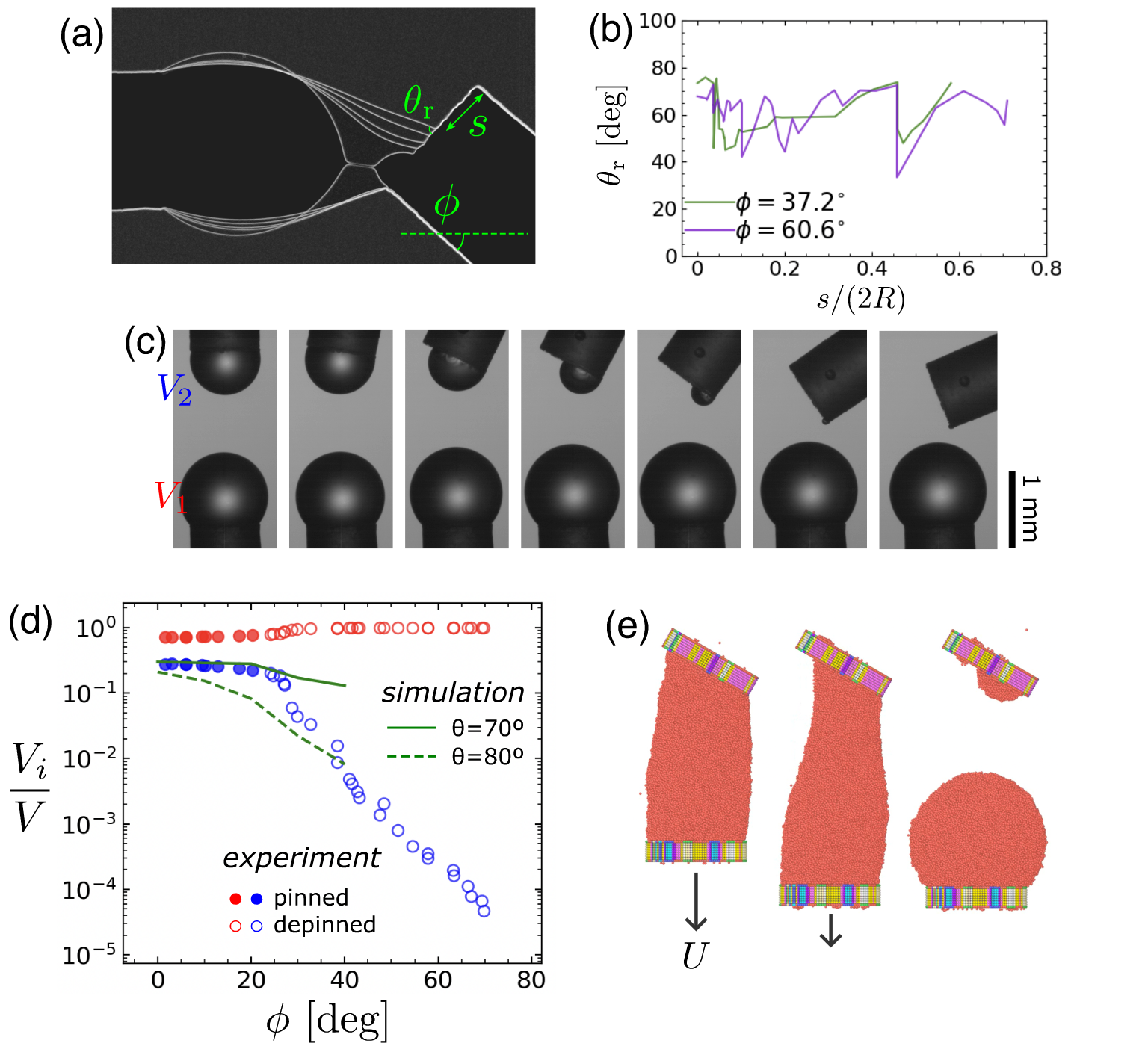}
	\caption{
    Depinning the contact line allows a larger range of splitting ratios.
	(a) Image overlays of the profiles of a liquid bridge between the tips of two VPS polymer fibers at an inclination $\phi = 49.6^{\circ}$.
    (b) The receding contact angle $\theta\ts{r}$ as the contact line slides across fiber face, for two rod angles.
    (c) Post-rupture states between two VPS fibers with a range of $\phi$.  
	(d) Post-rupture volumes versus $\phi$.  The smaller droplet reduces its size by 4 orders of magnitude as $\phi$ increases, facilitating a near-complete transfer.  Solid markers: pinned contact line at low $\phi$.  Open markers: depinned contact line at large $\phi$. Solid and dashed lines: MDPD simulations for two effective contact angles $\theta$.
    (e) Snapshots of the MDPD simulation with $\theta=80^{\circ}$, exhibiting contact-line depinning.
	}
	\label{fig:depin}
\end{figure}

\subsection{Complete droplet transfer}
If the fluid can be coaxed into contacting a smaller area of the tilted rod, it may have the effect of mimicking a smaller rod that would collect a smaller share of the liquid. 
To test this idea, we perform droplet-splitting experiments using vinylpolysiloxane (VPS) fibers (Zhermack 8 shore A) of radius 540 $\mu$m, which we make by pouring mixed precursor fluids into custom molds and letting them cure, following Ref.~\cite{nasto16}. We produce a clean, well-controlled fiber face by creating an excess thin film of VPS (much larger than the fiber face), which is peeled off to reveal a flat and smooth fiber face (see SI). Since the residue droplets now can be much smaller, we use a water/glycerol mixture (viscosity $\eta=104$ cP) to reduce evaporation.  Figure~\ref{fig:depin}a shows the droplet-splitting process 
now is complex: the liquid contact line may depin and evolve simultaneously with the shape of the liquid bridge as the rods are pulled apart. The receding contact angle, $\theta_\ts{r}$, is measured to be around $70^{\circ}$, shown in Fig.~\ref{fig:depin}b.
A substantial variation of the contact angle with the contact-line displacement, $s$, reveals a significant stick-slip motion.  Before the contact line sweeps across the entire surface, the liquid bridge becomes unstable and pinches off from the bridge neck. 

Despite this complexity, the end-result gives a simple trend. Figure~\ref{fig:depin}c shows the end states of a series of experiments splitting a liquid bridge of $V=1.45$ L between two VPS fibers at a sequentially larger angle $\phi$, where the residue volume remaining in the top fiber decreases drastically at large angles.
We measure the final droplet volumes on each fiber and show in Fig.~\ref{fig:depin}d that they are relatively flat up to $\phi \approx 25^{\circ}$ when the contact line remains pinned. For $\phi \gtrsim 25^{\circ}$, the contact line evolves before pinch-off, leading to a sharp decline in the residue $V_2$ by 4 orders of magnitude. At $\phi\approx70^{\circ}$, over 99.995\% of the liquid transfers to the bottom fiber. As $\phi$ increases further, the liquid bridge starts to creep onto the side wall, leading to a slightly increased post-rupture residue. Here, we focus on the regime $\phi\lesssim 70^{\circ}$ where this complication does not occur.

\begin{figure}[tb]
	\centering
	\includegraphics[width=0.95\textwidth]{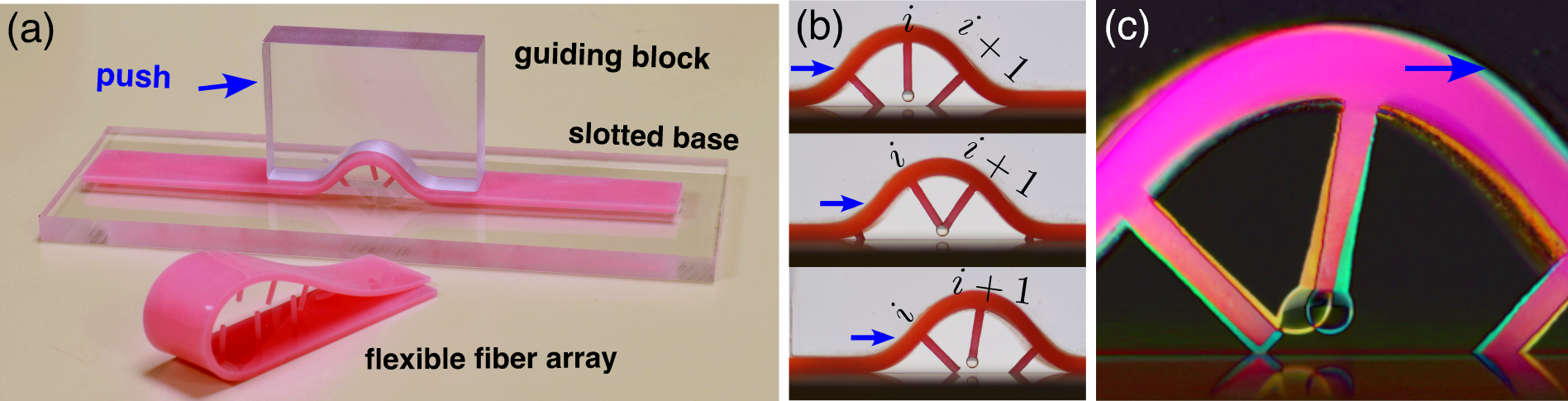}
	\caption{
	Using material deformations to move a droplet from fiber tip to fiber tip. (a) A VPS elastomer is molded into a flexible fiber array.  A guiding block imposes a ruck of the flexible substrate with a designed angles between the adjacent fibers.  (b) Demonstration of a sustained droplet transfer, when the guiding block is continuously pushed in one direction (blue arrows).  A drop is nearly completely transferred from site $i$ to $i+1$.  (c) Subtraction of two consecutive frames at the moment of liquid bridge pinch-off.
	}
	\label{fig:fiberarray}
\end{figure}

\subsection{Mesoscale simulations} 
To further explore the effect of the contact angle on the post-rupture volumes, we use mesoscale simulations using the Many-body Dissipative Particle Dynamics (MDPD) method, since it is much more convenient to tune the contact angle precisely~\cite{li2013three,chen2022coupling} in these simulations by adjusting the liquid-solid interactions.
This method is well suited for processes dominated by liquid viscosity and surface tension~\cite{warren2003vapor,arienti2011many}, ranging from droplet motions on rigid solid substrates~\cite{li2013three}, dynamic capillary wetting~\cite{cupelli2008dynamic}, self-cleaning of hydrophobic rough surfaces~\cite{zhang2019self}, and droplet wrapping of an elastic rod~\cite{chen2022coupling}. 
The simulations follow the same setup as in experiments with a liquid bridge (1.45 $\mu$L) placed between two rods at different inclined angles, whose radii are close to 500 $\mu$m. The liquid has an effective surface tension of $21.2$ mN/m and kinetic viscosity of $2.9 \times 10^{-5}$ m$^2$/s. 
The rods are discretized into particle clouds and the bottom rod is assigned a uniform velocity to mimic the rigid-body motion. 
To compare with our experiments given the measured contact angles (Fig.~\ref{fig:depin}b), we ran our simulation at two contact angles, $\theta=70^{\circ}$ and $\theta=80^{\circ}$, and the resulting volume ratios are shown in Fig.~\ref{fig:depin}d (solid and dashed lines).
Note that the MDPD method yields an effective contact angle with negligible hysteresis, which is difficult to achieve in experiments. 
Despite this simplification, the simulations capture the experimentally observed trend of $V_2/V$ as a function of $\phi$, as well as the depinning process of the contact line (Fig.~\ref{fig:depin}e).

\subsection{Pinned versus sliding contact line}
We note that so far, there are two possible scenarios for the contact line: (i) it remains pinned (for water on metal rods) (ii) it depins from the far edge to slide across the fiber face (for water/glycerol on VPS). 
The contact-line configuration generally depends on multiple scales of the triple-phase interface, involving solid curvature, surface chemistry, physical defects and the history of its motion - a debated subject not fully understood to date~\cite{bonn09, tadmor2021,wang2024,lindeman2024}. However, different from a liquid spanning extended objects~\cite{bush2010,bradley2019,butler2020,bradley2021, shao2024}, 
in our experiments, the geometrically sharp corners of the rods' faces help facilitate a range of contact angles through the Gibbs condition~\cite{gibbs1906, oliver1977}.  This allows us to treat the observed contact angles as the \emph{outcome} of force balance on the bridge, rather than  a boundary condition.  
We can thus target either pinning or depinning by picking appropriate rods  (e.g., metal or VPS, respectively). 
Accordingly, without the need to design the magnitude of contact angles, we have lumped the effects of the contact angles and their hysteresis into contact-line pinning (Figs.~\ref{fig:volume}-\ref{fig:pitchforkbif}) or depinning (Fig.~\ref{fig:depin}), which we have harnessed to the effect of droplet transfer.



\subsection{``Caterpillarity''}
The vanishing residue droplet on the VPS fiber face at large angles (Fig.~\ref{fig:depin}b) suggests that a single droplet can be passed along a series of fibers, so long as the motions and angles of the fibers are well-controlled. To demonstrate this concept, we mold a VPS elastomer to form a hairy surface that is elastically deformable. The cured sample is laid down on a base plate with a slot where the fibers hang freely. We form a ruck in the sample, and use a block to guide the ruck along the length of the sample, as shown by Fig.~\ref{fig:fiberarray}a. The shape of the bump in the guiding block as well as the length and the spacing of the fibers are designed to bring pairs of adjacent fibers toward each other, one by one, along the underside of the sample.

To operate the device, we attach a droplet of glycerol/water mixture ($\eta=104$ cP) to the end of one fiber $i$ (Fig.~\ref{fig:fiberarray}b, first panel). The propagating bump rotates this fiber counterclockwise, bringing the droplet into contact with the neighboring fiber $i+1$ to its right (second panel). As the bump passes through, fiber $i$ rotates outward again clockwise with a large angle, leaving the droplet almost entirely on the next fiber (last panel). Figure~\ref{fig:fiberarray}c illustrates the pinch-off by showing the difference between two subsequent images. Crucially, the orientation and motion of the rods create the conditions for near-complete transfer from Fig.~\ref{fig:depin}b.  In this way, the drop is transferred by one fiber in the traveling direction of the wave, so that it can repeat this process again. We show the entire process in the SI (movie).

\section{Conclusion}
It has been established since Savart~\cite{savart} that a \emph{free} liquid cylinder breaks due to an innate dynamic instability, insensitive to nozzle conditions~\cite{eggers1997, eggers2008}. In contrast, our angled two-rod system provides a route for external control over the end state of fragmentation.  By simply turning the relative angle of the rods, a pair of binary post-rupture states can be accessed precisely.  
For the case of a pinned contact line, our experiments and analysis show that the transition between the post-rupture states is essentially independent of the liquid properties (density, viscosity, and surface tension), and is thus applicable for any liquid to the first approximation.

Delving deeper, a tilt in our opposing rods is linked to the creation of distinct turning points in an otherwise perfect pitchfork bifurcation.  The robustness and sensitivity of our results are ultimately rooted in a \emph{connectivity change} in the bifurcation diagram (Fig.~\ref{fig:pitchforkbif}).
Unlike other biasing effects from the axial flow, gravity or density~\cite{atreya2002,kumar2015}, the angle change, combined with contact line motions, provides an easily tunable, \emph{geometric} handle that should apply at microscopic scales, where body forces fall to the wayside as surface forces become dominant.

Soft materials designed with a repeated unit cell can exhibit remarkable mechanical responses not inherent to their native materials~\cite{bertoldi2017}. 
Yet, the potential of designing mechanical metamaterials that can manipulate fluids 
are avenues that remain relatively unexplored \cite{li2021}.
Here we attain a high degree of control over a liquid droplet with an otherwise unremarkable polymer surface, by harnessing the material geometry to transfer liquid from one pillar to another. Our device may serve as a proof-of-concept for the broader idea of using patterned solids to drive liquids and vice versa.

\section{Acknowledgements}
We thank Marko Suchy for helpful discussions, and M.H. thanks Philip Frank Arnold and David Pratt for their assistance in preparing the experimental apparatus. 
This work was supported by NSF Awards REU-DMR-1757749 (M.M.), DMR-CAREER-1654102 (M.H., J.D.P), DMR-2318680 (J.D.P.), CMMI-CAREER-1847149 (H.J., T.Z.), EFRI-1935294 (S.H., C.D.S.), a Syracuse University BioInspired Seed Grant (M.H., S.H.), and Syracuse University research subsidy funds (C.D.S.). 
The technology described in this manuscript is covered by pending US patent application number US 63/786,596~\cite{weta_patent}.
M.H. and S.H. contributed equally to this work.

%

\clearpage

\renewcommand{\thefigure}{S\arabic{figure}}
\setcounter{figure}{0} 
\renewcommand{\theequation}{S\arabic{equation}}
\setcounter{equation}{0}
\renewcommand{\thesection}{SI \arabic{section}}
\setcounter{section}{0} 

\noindent\textbf{\LARGE Supplementary Information for \\} \\
\noindent\textbf{\Large ``Controlling Droplets at the Tips of Fibers" \\} \\
\noindent\textbf{\footnotesize Mengfei He, Samay Hulikal, Marianna Marquardt, Hao Jiang,Anupam Pandey, Teng Zhang, Christian D. Santangelo and Joseph D. Paulsen}

\section{Angled liquid bridge: Experiments}
\subsection{Experimental method}
\begin{figure}[h]
\includegraphics[width=0.6\textwidth]{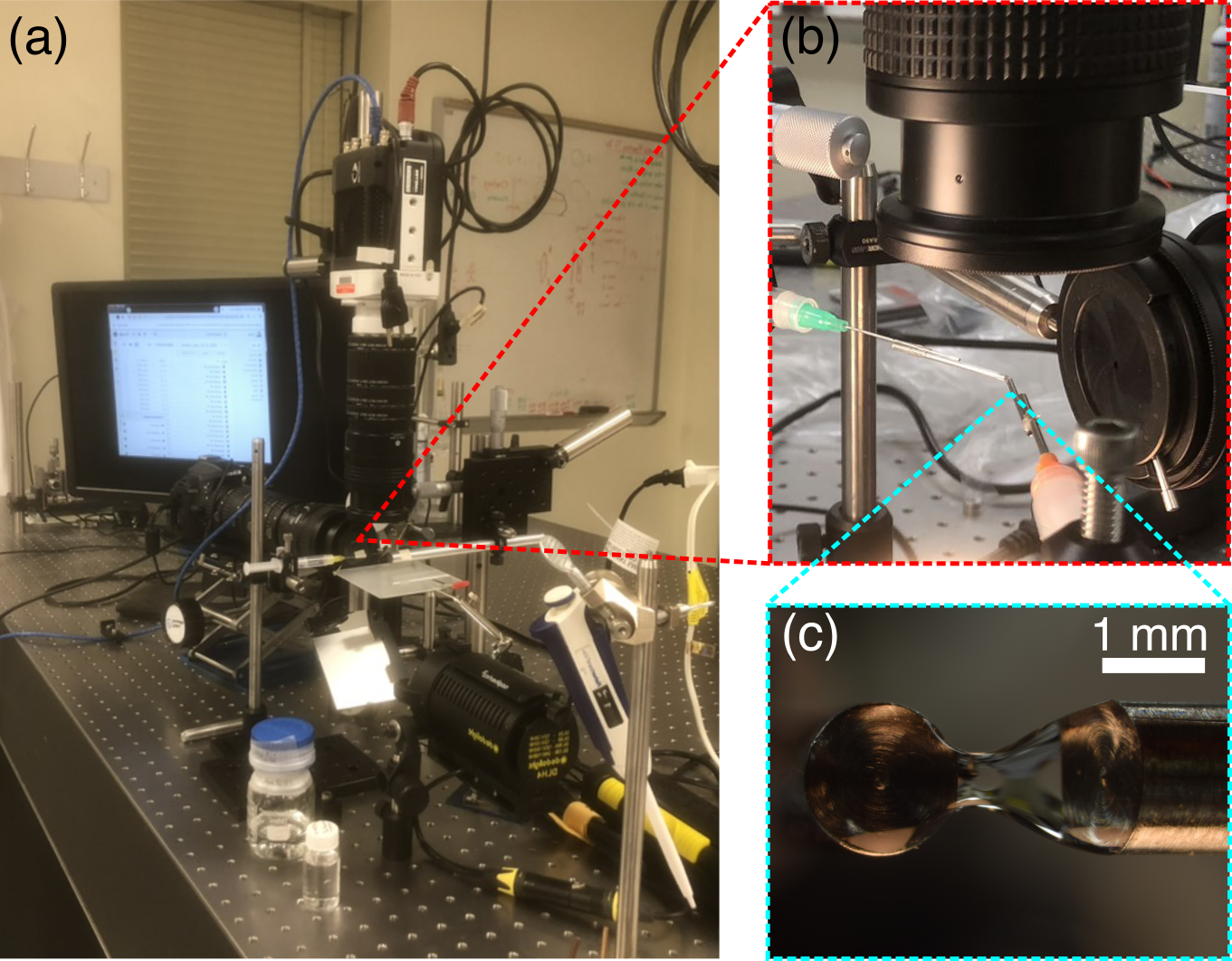}
\caption{
a) Experimental setup.  A droplet is gapped by two micro rods, viewed from above with a high-speed camera and from the side with a DSLR camera.
b) The angled rods are oriented horizontally to minimize the biasing effect of gravity.  An external pinhole is applied to the side-view camera to increase the depth of field.
c) The photograph captured by the side-view camera.
}
\label{fig:setup}
\end{figure}

Our setup is shown in Fig.~\ref{fig:setup}a. To form a liquid bridge and affix its contact lines, we pipette a droplet of de-ionized water (18.4 M$\Omega$ cm, Milli-Q IQ 7005) and place it between two steel gage pins (Vermont Gage A series, diameter $0.14 < R < 0.76$ mm).
We first tilt one rod by an angle $\phi$, and then slowly enlarge the spacing between the two rods by displacing the nontilted rod along its axial direction while maintaining the relative angle $\phi$ (Fig.~\ref{fig:setup}b).
The rupture process of the liquid bridges is recorded from the top by a high-speed camera (Phantom Miro Lab 340) connected to a macro lens (Nikkor 60 mm F/2.8, 80 mm extension tube).  We use a second camera (DLSR, Nikon D5300 with the same lens and extension tube) to monitor the horizontal alignment of the two rods (Fig.~\ref{fig:setup}a, b).
The large contact angle hysteresis and the sharpness of the rod edges pin the contact lines of the liquid bridge during the elongation and pinch-off processes (Fig.~\ref{fig:setup}c). 

To test the effect of the droplet viscosity, glycerol is mixed with de-ionized water to obtain sample liquid of viscosity $1 < \eta < 104$ cP, measured by a rheometer (Anton Paar MCR302).  Over two decades of the viscosity change, we do not find the post-rupture droplet sizes (presented in the main text) to be altered by more than $\pm2\%$.

\subsection{Axisymmetry of a tilted liquid bridge}
\begin{figure}[htp]
\includegraphics[width=0.7\textwidth]{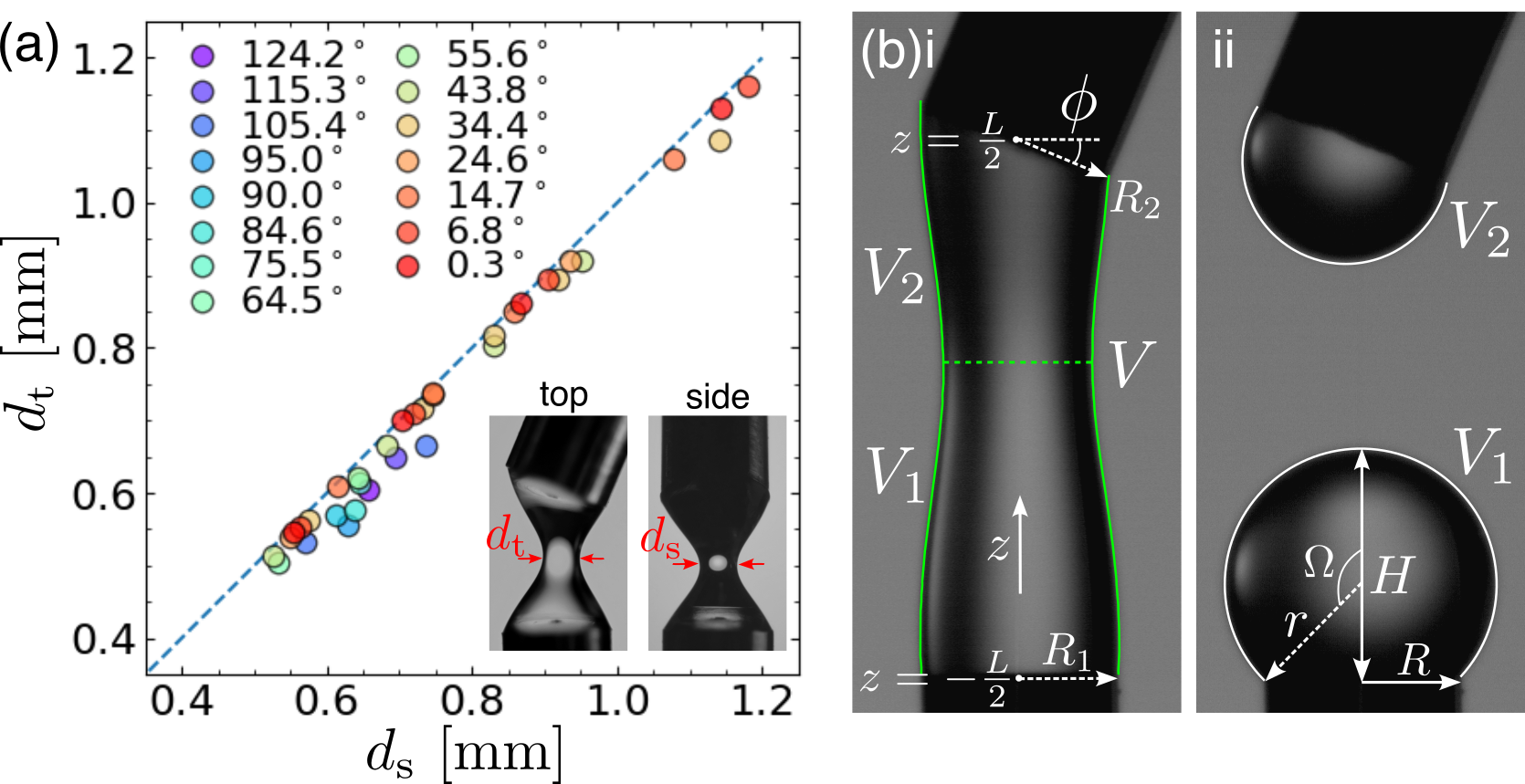}
\caption{
a) Neck size of the liquid bridges measured from the top ($d\ts{t}$) and from the side ($d\ts{s}$) match well up till near $90^{\circ}$ of the rod inclination, suggesting approximate axisymmetry of the tilted liquid bridge.
b)i Top view of a tilted liquid bridge, with its profile captured by an edge-finding algorithm (green curves).  Dashed line: location of the minimum bridge width from the edge-finder, defined as the neck. ii Spherical caps (white curves) are manually fitted to the post-rupture droplets to calculate the separate volumes.
}
\label{fig:sym}
\end{figure}

Far from the inclined boundaries of a tilted liquid bridge, axisymmetry of the bulk liquid bridge is observed to be partially preserved.  To test this, we measure the neck size of a liquid bridge from the top-view ($d\ts{t}$) and the side-view ($d\ts{s}$) simultaneously.  Figure~\ref{fig:sym}a shows that $d\ts{s}$ and $d\ts{t}$ from the two orthogonal views are in close proximity to each other, even when the inclination is close to $90^{\circ}$.

The approximate axisymmetry of a liquid bridge, from the nontilted boundary up to the bridge neck, allows us to calculate its volume partition by its neck.  To do so, we track the profile $F(z)$ of a liquid bridge along its axis ($z$) by thresholding (\textit{thresh\_binary, OpenCV}) a grayscale frame (solid lines, Fig.~\ref{fig:sym}b(i)).  We then calculate the volume from the nontilted side $V_1 = \int \pi F(z)^2\,dz$, where the integral is bounded by the nontilted boundary and the neck (dashed line, Fig.~\ref{fig:sym}b(i)).  When the tilt angle of the far-end rod is large, there is a slight inclination of the overall bridge axis.  The estimation of $V_1$ is robust to this slight overall inclination, in which case we simply integrate along the vertical distance between the nontilted rod and the plane of the neck.  Given the total volume of the liquid bridge $V$, the partitioned proportions can be obtained as $V_1/V$ and $(V-V_1)/V$.
As the gap of the two rods increases, the liquid bridge eventually ruptures into two separate droplets attached to the rod faces, as shown in Fig.~\ref{fig:sym}b(ii).  We manually fit spherical caps to these droplets and calculate their volumes by $V_i=\pi r_i^3(2+\cos\Omega_i)(1-\cos\Omega_i)^2/3$, where $i=1$ and $2$ correspond to the top and bottom droplets in the image, respectively.  When a droplet is shallower than a hemisphere $\Omega<\pi/2$, we use the alternative formula  $V_i = \pi H_i (3R_i^2+H_i^2)/6$, since measuring the cap height $H$ is more precise than fitting a small part of a sphere.

\subsection{Polymeric fiber array}
\begin{figure}[htp]
\includegraphics[width=0.7\textwidth]{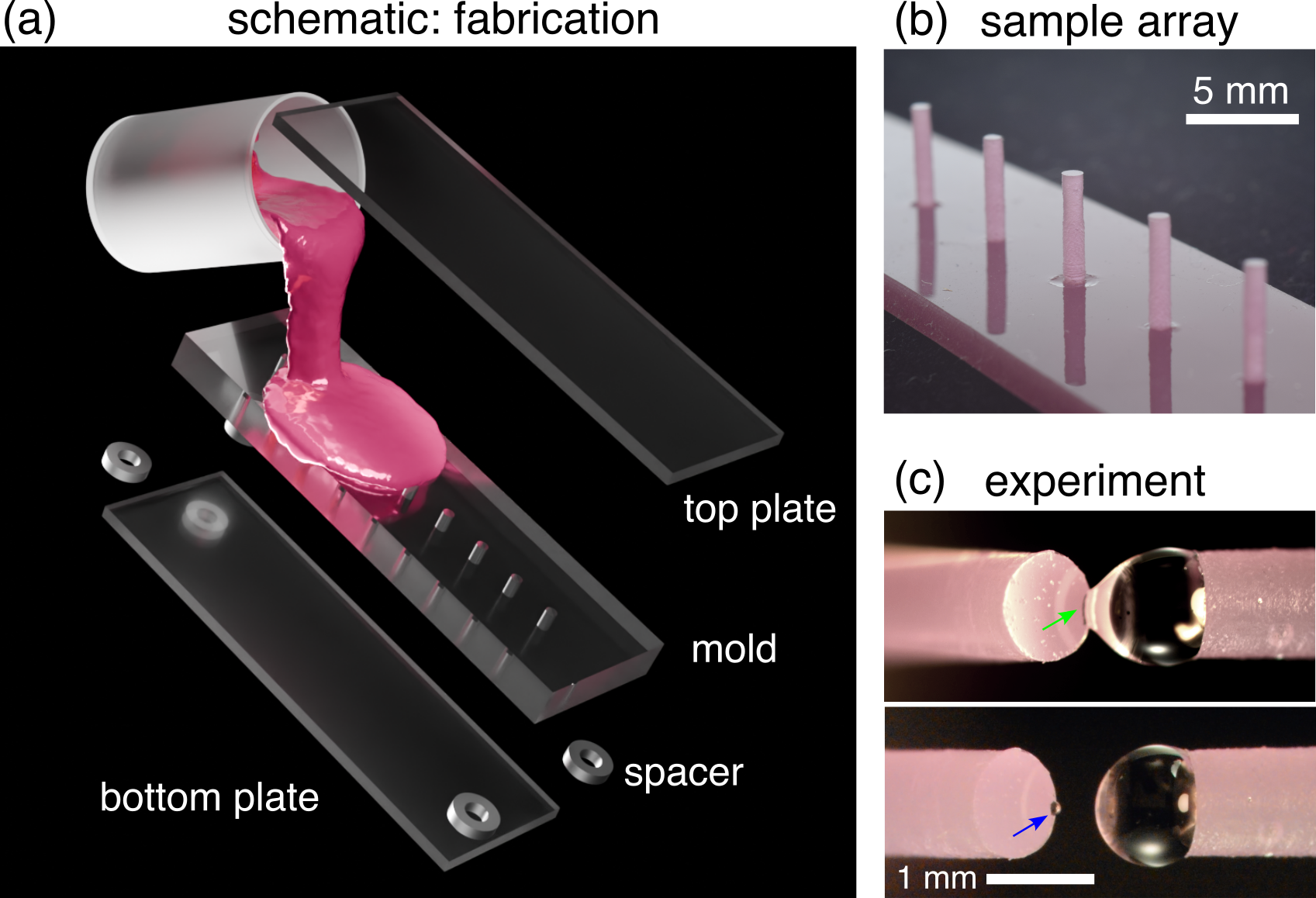}
\caption{
a) Schematic: Vinylopolysiloxane (VPS-8) is poured onto an acrylic mold, clamped by the top and bottom plates,  to cure into a polymeric fiber array.
The spacers between the mold and the bottom plate set the thickness of the polymeric substrate.
b) A completed sample fiber array. 
c) The contact line of a liquid bridge depins and slides across the tilted (left) fiber face.  A receding contact line (green arrow) gives rise to a small residue droplet (blue arrow) as a result of the liquid bridge pinch-off.
}
\label{fig:vps}
\end{figure}

A schematic of the fabrication of a polymeric fiber array is shown in Fig.~\ref{fig:vps}a.  We first mix Vinylpolysiloxane (VPS, Zhermack Elite Double 8 shore A) base and cure agents with a weight ratio of 1:1. The mixture is placed in vacuum for 4 minutes at room temperature (25$^{\circ}$C) to remove micro bubbles.  The precursor fluid is then poured onto an acrylic mold drilled with an array of through-holes.  After the mixture fluid perfuses through the array of holes, we clamp the top and bottom plates onto the mold, allowing the fluid mixture to cure (20 minutes) into an array of fibers with a planar substrate.  An extremely thin VPS film forms between the mold and the top plate.  We carefully peel off this excess film after curing, to produce a flat, smooth fiber face that allows the liquid contact line to depin and slide across. The thickness of the acrylic mold sets the length of the fibers, and the thin spacers between the bottom plate and the mold set the thickness, and thus the flexibility, of the VPS substrate.  The cured elastomer is lastly demolded from the bottom side.  A completed sample is shown in Fig.~\ref{fig:vps}b.

Figure~\ref{fig:vps}c shows a close view of the interaction of two polymeric fibers with a single droplet, captured horizontally from the side using a DSLR (setup configuration shown in Fig.~\ref{fig:setup}a).  The left fiber is tilted (in the horizontal plane) with respect to the line connecting the centers of the two fiber faces.  As the gap between two fibers widens, the contact line depins from the left side, sliding across the face of the left fiber (green arrow).  With the solid-liquid contact area shrinking, the left-right asymmetry of the liquid bridge is greatly amplified. The liquid bridge eventually pinches off from the bridge neck, leaving a residual droplet on the left (blue arrow).

\section{Angled liquid bridge: Theory}
The lack of axial symmetry of an angled liquid bridge requires us to confront a non-linear partial differential equation with complicated boundary conditions. 
Although numerical techniques have been employed to study liquid bridges under this lack of symmetry \cite{chen1992s}, our primary mathematical apparatus will be the Lyapunov-Schmidt procedure for one-dimensional kernels, as detailed in Refs. \cite{textbooks} and \cite{lowgravs}.
For liquid bridges, the procedure has been utilized to analyze the stability of bridges that are supported between parallel, almost circular contacts \cite{meseguer2001s}, between unequal circular contacts and subjected to an axial acceleration \cite{meseguer2003s}, and bridges involving fluid flow and gravity \cite{atreya2002s}.

We assume that the fluid flow generated when the bridge starts to collapse occurs for such a short period of time that negligible fluid moves across the narrowest section (referred to as the neck) of the bridge as it breaks.
This implies that the dynamics will preserve any asymmetry across the static bridge’s neck at its stability limit. 
Therefore, to determine the threshold angle, we need to find the tilt at which the shape of the capillary bridge, at its stability limit, is as symmetric as possible on both sides of the neck.

\subsection{Equilibrium equation, boundary conditions, constraints}\label{second}

Figure~\ref{fig:sym}b(i) depicts a capillary bridge held between circular supports with the rod that has the larger radius, $R_2$, tilted by an angle $\phi$. 

We define a cylindrical coordinate system by orienting the $z$-axis along the line joining the centers of the circular rods.
The coordinate along this axis is denoted by $z$.
The angled circular support is centered at $z=+L/{2}$ and is tilted by an angle $\phi$.
The level rod has a radius $R_1$ and is centered at $z=-L/2$.
The cross-section of the bridge at $z$ has a shape $F(z,\theta)$.
We will use the length scale $R_0=(R_1+R_2)/{2}$ to construct dimensionless parameters:
\begin{equation}
    \Lambda=\frac{L}{2 R_0},\quad h= \frac{R_2-R_1}{R_2+R_1},\quad \mathcal{V}= \frac{V}{R_0^{3}},
\end{equation}
where $\Lambda, h$, and $\mathcal{V}$ are the dimensionless length, dimensionless difference in rod radii, and the dimensionless volume of the bridge respectively. 
We also define a dimensionless coordinate, $x$, and a dimensionless cross-sectional shape, $f(x,\theta)$, of the bridge as
\begin{equation}
    x=2\pi \frac{z}{L},\quad
    f(x,\theta)=\frac{1}{R_0} F\left(z=\frac{L}{2\pi}~x,\theta\right).
\end{equation}

\subsubsection{Equilibrium equation}
A general capillary surface has constant mean curvature at every point \cite{lowgravs}. 
With the shape of the bridge specified by $f(x,\theta)$, this condition is given by the non-linear partial differential equation
\begin{equation}\label{eq:mce}
    M[\Lambda,f(x,\theta)]+Q=0,
\end{equation}
where $M/2$ is the dimensionless mean curvature of the surface at $(x,\theta)$, given by
\begin{multline}
    M \left[\Lambda, f(x,\theta) \right] =
    \Lambda\left[ 
    \Lambda^{2}\partial_{\theta}f^{2}+f^{2}(\Lambda^{2}+\pi^{2}\partial_{x}f^{2})
    \right]^{-3/2} 
    \Big[
    f(\Lambda^{2}+
    \pi^{2}\partial_{x}f^{2})(\partial_{\theta}^{2}f-f)\\
    +\pi^{2}f~\partial_{x}^{2}f(f^{2}+\partial_{\theta}f^{2})-
    2\partial_{\theta}f(\Lambda^{2}\partial_{\theta}f+\pi^{2}f\partial_{x}f\partial_{x}\partial_{\theta}f)
    \Big],
\end{multline}
where the derivatives are defined as $\partial_x\equiv\frac{\partial}{\partial x}$ and $\partial_\theta\equiv\frac{\partial}{\partial \theta}$.
$Q$ is an undetermined constant that incorporates the surface tension and pressure difference between the fluid and the surrounding medium.
We will refer to $Q$ as the dimensionless pressure. 

\subsubsection{Boundary conditions at circular supports}
The level rod has the boundary condition
\begin{equation}\label{eq:bbd}
    f\left( 
-\pi,\theta
\right)=1-h.
\end{equation}
The tilted support extends from $z=L/2-R_2\sin \p$ at its lowest point to $z=L/2+R_2\sin \p$ at its apex.
The boundary condition for this rod is
\begin{equation}\label{eq:topbd}
    f\left( 
\pi- \frac{\pi(1+h)}{\Lambda} \frac{\tan \phi \cos\theta}{\sqrt{ 1+\tan ^{2}\phi \cos ^{2}\theta }},\theta
\right)=
\frac{1+h}{\sqrt{ 1+\tan ^{2}\phi\cos ^{2}\theta }}.
\end{equation}

\subsubsection{Volume constraint}
The dimensionless volume contained inside the liquid bridge, $f(x,\theta)$, is given by
\begin{multline}\label{eq:vc}
    \mathcal{V}= \frac{\Lambda}{2\pi}\int _{-\pi}^{+\pi}\int _{-\pi}^{+\pi}f(x,\theta)^{2} \, d\theta  \, dx
+ \int _{+\pi /2}^{+\pi}\int _{-\pi}^{+\pi}\frac{(1+h)\sin \phi \sin \beta}{2}
\Big[
f\left( \pi- \frac{\pi(1+h)}{\Lambda}\sin \phi \cos\beta,\theta \right)^{2} \\
-f\left( \pi- \frac{\pi(1+h)}{\Lambda}\sin \phi \cos\beta, \left( \frac{\pi-\alpha}{\pi} \right)\theta  \right)^{2} \left( \frac{\pi-\alpha}{\pi} \right)
-f\left( \pi+ \frac{\pi(1+h)}{\Lambda}\sin \phi \cos\beta, \frac{\alpha}{\pi}\theta \right)^{2} \left( \frac{\alpha}{\pi} \right) \Big]   \, d\theta  \, d\beta ,
\end{multline}
where $\alpha$ is defined as
\begin{equation}\label{eq:alp}
    \cos\alpha =- \frac{\cos \phi \cos\beta}{\sqrt{ 1-\sin ^{2}\phi \cos ^{2} \beta }}.
\end{equation}
The complexity arises due to the fact that there is fluid only on one side of the rod from $z=L/2-R_2\sin \p$ to $z=L/2+R_2\sin \p$.
Notice that we have the dimensionless shape $f$ extending beyond $x=+\pi$ at the tilted rod in the exact expressions for its boundary condition (Eq.  \ref{eq:topbd}) and the volume constraint (Eq. \ref{eq:vc}).
However, the perturbation expansion will transform them into conditions on the derivatives of $f$ at $x=+\pi$ and add corrections to the volume integral of the bridge up to $x=+\pi$.


\subsection{Lyapunov-Schmidt decomposition for angled liquid bridges}\label{third}
We choose a cylindrical bridge as the base shape for the perturbation. 
Such a bridge is described by the shape $f(x,\theta)=1$ and is obtained when the circular supports are parallel ($\p=0$) and have the same radii ($h=0$).
Moreover, we pick the cylinder's length to be at its stability limit \cite{mesegs}, given by $\Lambda=\pi$.
The volume of the cylinder is $\mathcal{V}=2\pi^2$.
It is readily verified that these cylindrical parameter values and shape solve the differential equation (Eq. \ref{eq:mce}), with $Q=1$, while satisfying the boundary conditions (Eqs. \ref{eq:bbd}, \ref{eq:topbd}) and volume constraint (Eq. \ref{eq:vc}).

We define the parameter space, $X$, and the shape space, $Y$, as small deviations from a cylindrical bridge:
\begin{equation}
    X=(\lambda,h,\phi,v),\quad
    Y=(q,f(x,\theta)),
\end{equation}
where $\lambda$, $h$, $\phi$, and $2\pi^2\cdot v$ are deviations of the length, rod radii, tilt angle, and volume parameters, respectively, while $q$ and $f(x,\theta)$ are deviations in the dimensionless pressure and shape, respectively. 
Next, we construct a function $\mathcal{F}$ that maps $X\times Y$ to a codomain space, $\mathcal{P}$, as
\begin{equation}
    \mathcal{F}((\lambda,h,\phi,v),(q,f))=
    \Big( F_{1}(x,\theta), F_{2}(\theta),F_{3}(\theta),F_{4} \Big),
\end{equation}
where $F_{1}$ encodes the extremum equation (Eq. \ref{eq:mce}):
\begin{equation}
    F_{1}(x,\theta)=M[\pi+\lambda,1+f(x,\theta)]+(1+q),
\end{equation}
$F_{2}$ and $F_3$ involve the boundary condition at the level and tilted rods, respectively (Eq. \ref{eq:bbd}, \ref{eq:topbd}):
\begin{align}
     F_{2}(\theta) & = (1+f)\left( -\pi,\theta\right)-(1-h),\\
    F_{3}(\theta) & = (1+f)\left( 
\pi- \frac{\pi(1+h)}{\pi+\lambda} \frac{\tan \p \cos\theta}{\sqrt{ 1+\tan ^{2} \p \cos ^{2}\theta }},\theta
\right)-
\frac{1+h}{\sqrt{ 1+\tan ^{2}\p\cos ^{2}\theta }},
\end{align}
and $F_{4}$ has the volume constraint (Eq. \ref{eq:vc}):
\begin{multline}
F_{4}= -\left(  1 +v \right) + \frac{\pi+\lambda}{4\pi^{3}} \int _{-\pi}^{+\pi}\int _{-\pi}^{+\pi} (1+f)(x,\theta)^{2} \, d\theta  \, dx \\
+ \int _{+\pi /2}^{+\pi}\int _{-\pi}^{+\pi}\frac{(1+h)\sin\p  \sin \beta}{4\pi^2}
\Big[ 
(1+f)\left( \pi- \frac{\pi(1+h)}{\pi+\lambda}\sin \p \cos\beta,\theta \right)^{2}  \\
-(1+f)\left( \pi- \frac{\pi(1+h)}{\pi+\lambda}\sin \p \cos\beta, \left( \frac{\pi-\alpha}{\pi} \right)\theta  \right)^{2} \left( \frac{\pi-\alpha}{\pi} \right)\\
-(1+f)\left( \pi+ \frac{\pi(1+h)}{\pi+\lambda}\sin \p \cos\beta, \frac{\alpha}{\pi}\theta \right)^{2} \left( \frac{\alpha}{\pi} \right) \Big]   \, d\theta  \, d\beta,
\end{multline}
with $\alpha$ defined in Eq. \ref{eq:alp}.
Given the deviations of the parameters, $(\lambda, h, \phi, v)$, the shape $(q,f(x,\theta))$ is obtained from the bridge equation:
\begin{equation}\label{eq:be}
    \mathcal{F}((\lambda,h,\phi,v),(q,f))=0.
\end{equation}
The cylindrical bridge satisfies $\mathcal{F}((0,0,0,0),(0,0))=0$.

The Fréchet derivative \cite{textbooks} of $\mathcal{F}$ is a linear map operating on a shape $(q,f)$.
When evaluated at the cylinder, it is given by
\begin{equation}\label{eq:frechet}
d_{y}\mathcal{F}(q,f)=
\Big(\nabla^{2}f+f+q, f(-\pi,\theta), f(+\pi,\theta), \frac{1}{4\pi^{2}}\int _{-\pi}^{+\pi}\int _{-\pi}^{+\pi}2~f(x,\theta) \, d\theta  \, dx  \Big),
\end{equation}
where $\nabla^{2}\equiv \partial_{x}^{2}+\partial_{\theta}^{2}$ is the Laplacian operator.
The map $d_{y}\mathcal{F}$ and its adjoint, $d_{y}\mathcal{F}^\dagger$, have one-dimensional kernels:
\begin{equation}
    \text{ker }d_{y}\mathcal{F}=\text{span}(u),\quad \text{ker }d_{y}\mathcal{F}^{\dagger}=\text{span}( s ),
\end{equation}
where $u=(0,\sin x)$ and $s= \frac{1}{\sqrt{ 2\pi^{2}+2 }}(2\pi \sin x,+1,-1,0)$. 
We use the kernels to split the spaces $Y$ and $\mathcal{P}$ as
\begin{equation*}
    Y=U\oplus U^{\perp},\quad\mathcal{P}=S\oplus S^{\perp},
\end{equation*}
where $\perp$ denotes the orthogonal complement of a subspace, $U=\text{span}(u)$, and $S=\text{span}(s)$. 
Defining $H$ as the projection operator onto $S$ and $\mathbb{I}$ as the identity operator on $\mathcal{P}$, we decompose the bridge equation (Eq. \ref{eq:be}) into two equations:
\begin{eqnarray}
\label{eq:lsshape}(\mathbb{I}-H)\mathcal{F}((\lambda,h,\phi,v),a(0,\sin x)+(q,w)) =0, &\\ 
\label{eq:lsbifur} H\mathcal{F}((\lambda,h,\phi,v),a(0,\sin x)+(q,w)) =0, &
\end{eqnarray}
where we have written the shape as $(q,f)=a(0,\sin x)+(q,w)$ with $a\in\mathbb{R}$, $a(0,\sin x)\in U$, and $(q,w)\in U^{\perp}$.
Through an inner product on $\mathcal{P}$ defined as
\begin{multline}\label{eq:ipP}
    \braket{ r | t }_{\mathcal{P}}=\frac{1}{4\pi^{2}}\int _{-\pi}^{\pi}\int _{-\pi}^{+\pi}r_{1}(x,\theta)t_{1}(x,\theta) \, dx  \, d\theta +
    \frac{1}{2\pi}\int_{-\pi}^{\pi} r_{2}(\theta)t_{2}(\theta) \, d\theta  +
    \frac{1}{2\pi}\int_{-\pi}^{\pi} r_{3}(\theta)t_{3}(\theta) \, d\theta+ r_{4}t_{4},
\end{multline}
for all $r\equiv(r_{1}(x,\theta),r_{2}(\theta),r_{3}(\theta),r_{4}),t\equiv(t_{1}(x,\theta),t_{2}(\theta),t_{3}(\theta),t_{4})\in \mathcal{P}$, the projection operator $H$ can be expressed as $H=s\braket{s | \cdot}_\mathcal{P}$.

The implicit function theorem \cite{textbooks} applies to Eq. \ref{eq:lsshape} and so, there exist unique functions $q(\lambda,h,\phi,v,a)$ and $w(x,\theta;\lambda,h,\phi,v,a)$ around $\lambda=0$, $h=0$, $\phi=0$, $v=0$ and $a=0$ such that Eq. \ref{eq:lsshape} is true. 
The real number $a$ is found using Eq. \ref{eq:lsbifur}.

Even with this decomposition, a direct attempt at an analytic solution is unfeasible. 
We expand $q$ and $w$ around $\lambda=h=\phi=v=0$ and $a=0$ as a power series to obtain
\begin{equation}\label{eq:beforeb}
    q=\sum_{i,j,k,l,m=0}^{\infty}\lambda^{i}h^{j}\phi^{k}v^{l}a^{m}q_{ijklm},\quad w =\sum_{i,j,k,l,m=0}^{\infty}\lambda^{i}h^{j}\phi^{k}v^{l}a^{m}w_{ijklm}(x,\theta),
\end{equation}
with $(q_{ijklm},w_{ijklm})\in U^{\perp}$ and $q_{00000}=0,w_{00000}(x,\theta)=0$. 
To reduce the number of parameters involved, we define a new placeholder variable $b$ as
\begin{equation}
    q  =\sum_{n,m=0}^{\infty}b^{n}a^{m}q_{nm},\quad
    w  =\sum_{n,m=0}^{\infty}b^{n}a^{m}w_{nm}(x,\theta),
\end{equation}
making $q_{nm}$ and $w_{nm}(x,\theta)$ homogeneous functions of the parameters $\lambda$, $h$, $\phi$, and $v$.
That is, for any real number $c$, we have
\begin{align}
q_{nm}(c\lambda,c h,c \phi,c v) & =c^{n} q_{nm}(\lambda,h,\phi,v) \\
w_{nm}(x,\theta;~c\lambda,c h,c \phi,c v) & =c^{n} w_{nm}(x,\theta;~\lambda, h,\phi, v).
\end{align}
We recover the expansion in Eq. \ref{eq:beforeb} upon replacing $b$ with $1$. 
Further, we let $\mathcal{F}=\sum_{n,m=0}^{\infty}b^{n}a^{m}\mathcal{F}_{nm}$.
Thus, Eqs. \ref{eq:lsshape} and \ref{eq:lsbifur} reduce to
\begin{align}
    \label{eq:ba1}(\mathbb{I}-H)\mathcal{F}_{nm} &= \mathcal{F}_{nm}-s\braket{ s | \mathcal{F}_{nm} } _{\mathcal{P}}  =0\\
    \label{eq:ba2}\sum_{n,m=0}^{\infty}b^{n}a^{m}\braket{ s | \mathcal{F}_{nm} } _{\mathcal{P}}&=\sum_{n,m=0}^{\infty}b^{n}a^{m}~\Omega_{nm} =0
\end{align}
where $\Omega_{nm}=\braket{ s | \mathcal{F}_{nm} }_{\mathcal{P}}$ and $H=s\braket{s | \cdot}_\mathcal{P}$.
The first of these equations determines the shape of the bridge as a power series:
\begin{equation}\label{eq:shapeseries}
    (1+f)(x,\theta)=1+a \sin x+\sum_{n,m=0}^{\infty}b^{n}a^{m}w_{nm}(x,\theta).
\end{equation}
Equation \ref{eq:ba2} is called the bifurcation equation, and it determines $a$ as a function of the parameters $\lambda$, $h$, $\phi$, and $v$. 


\subsection{Solving the bifurcation equation}\label{fourth}
To illustrate the typical steps used in finding $q_{nm}$, $w_{nm}$, and $\Omega_{nm}$, we consider $\mathcal{F}_{10}$.
We obtain $\mathcal{F}_{10}$ as
\begin{align}
    \mathcal{F}_{10} &=\Big(  
        \nabla^{2}w_{10}+w_{10}+q_{10},
        w_{10}(-\pi,\theta)+h,
        w_{10}(+\pi,\theta)-h,
        \frac{1}{4\pi^{2}}\int _{-\pi}^{\pi}\int _{-\pi}^{\pi}2 w_{10} \, d\theta  \, dx + \frac{\lambda}{\pi}-v\Big) \\
                    & = d_y\mathcal{F}\Big(q_{10},w_{10}(x,\theta)\Big)+(0,h,-h,\frac{\lambda}{\pi}-v),
\end{align}
with the simplification performed using the definition of the Fr\'echet derivative (Eq. \ref{eq:frechet}).
The space $S^\perp$ is equivalent to the image of $d_y\mathcal{F}$.
This property of the Fréchet derivative, along with $(q_{nm},w_{nm})\in U^\perp$, allows us to compute $\Omega_{nm}$.
We obtain $\Omega_{10}$ as
\begin{equation}
    \Omega_{10}=\braket{ s | \mathcal{F}_{10} }_{\mathcal{P}}= \frac{2h}{\sqrt{ 2\pi^{2}+2 }}. 
\end{equation}
The shape, $(q_{10}, w_{10})$, is found using Eq. \ref{eq:ba1},
\begin{equation}
    q_{10} =\frac{\lambda-\pi v}{2\pi},\quad
    w_{10}(x,\theta) =\frac{\pi v-\lambda}{2\pi}-\frac{\pi h}{\pi^{2}+1}x\cos x+\frac{\pi v-\lambda}{2\pi}\cos x-\frac{\pi h}{2(\pi^{2}+1)}\sin x.
\end{equation}
The perturbation terms are found sequentially.

We catalog two more terms in Eq. \ref{eq:shapeseries} that will be necessary for our purposes:
\begin{equation}
    w_{01}(x,\theta)=0,\quad
    w_{02}(x,\theta)=- \frac{1}{4}+ \frac{1}{4}\cos 2x.
\end{equation}
We will need to solve the bifurcation equation to find the maximum length of the bridge and the threshold angle. 
An inventory of the necessary $\Omega_{nm}$ in Eq. \ref{eq:ba2} is
\begin{multline}
    \Omega_{00}=0,\quad
    \Omega_{01}=0,\quad
    \Omega_{10}=\frac{2h}{\sqrt{ 2\pi^{2}+2 }},\quad  \Omega_{02}=0,\quad   \Omega_{11}=\frac{3\lambda-\pi v}{\sqrt{ 2\pi^{2}+2 }},\quad
    \Omega_{12}=-\frac{15 \pi ^2 h}{8 \sqrt{2} \left(1+\pi ^2\right)^{3/2}},\\
    \Omega_{20}=- \frac{8\pi h\lambda}{(2\pi^{2}+2)^{3 /2}}- \frac{\phi^{2}}{4\sqrt{ 2\pi^{2}+2 }},\quad
    \Omega_{03}=\frac{3\pi}{2\sqrt{ 2\pi^{2}+2 }},\\
    \Omega_{21}=-\phi^{2}\left( \frac{1}{4\pi \sqrt{ 2\pi^{2}+2 }} \right) 
 +h^{2}\left( \frac{\pi^{3}(63+8\pi^{2})}{16\sqrt{ 2 }~(1+\pi^{2})^{5/2}} \right)+v^{2}\left( \frac{11 \pi }{8 \sqrt{2\pi^{2}+2}} \right)\\
- \lambda^{2}\left(\frac{3 \left(7 \pi ^2-9\right)}{8\pi \sqrt{2} ~\left(1+\pi ^2\right)^{3/2}} \right)
- \lambda v
 \left( \frac{15+7 \pi ^2}{4 \sqrt{2}~ \left(1+\pi ^2\right)^{3/2}} \right),\quad\\
     \Omega_{30}=-h^{3}\left( 
\frac{\pi ^4 \left(97+24 \pi ^2\right)}{64 \sqrt{2} \left(1+\pi ^2\right)^{7/2}}
\right) +
h \lambda^{2}\left( 
\frac{95 \pi ^2-161}{32 \sqrt{2} \left(1+\pi ^2\right)^{5/2}}
\right)
 - 
\frac{v \phi^{2}}{8 \sqrt{2 \left(1+\pi ^2\right)}}
 +
\lambda \phi^{2}\left( 
\frac{1+5 \pi ^2}{8\pi \sqrt{2}   \left(1+\pi ^2\right)^{3/2}}
\right) \\
+
\frac{h \phi^{2}}{8 \sqrt{2} \left(1+\pi ^2\right)^{3/2}}
+h\lambda v\left( 
\frac{203 \pi }{48 \sqrt{2} \left(1+\pi ^2\right)^{3/2}}
\right)
-h v ^{2}\left( 
\frac{115 \pi ^2}{96 \sqrt{2} \left(1+\pi ^2\right)^{3/2}}
\right).
\end{multline}
We discard the placeholder variable $b$ in Eq. \ref{eq:ba2} and recover the complete bifurcation equation as
\begin{equation}\label{eq:bifeq}
    G(\lambda,h,\phi,v;a)\coloneq \sum_{n,m=0}^\infty a^{m}\Omega_{nm}(\lambda,h,\phi,v)=0.
\end{equation}

\subsubsection{Maximum length of angled bridges}
Bifurcation diagrams for angled liquid bridges are shown in Fig. 4 of the main text.
To find them, we numerically solve Eq. \ref{eq:bifeq} and plot the solutions as a function of length for two angled bridges with distinct tilt angles, as well as for a cylindrical bridge.
It is known that a cylindrical bridge breaks at a subcritical pitchfork bifurcation \cite{atreya2002s}. 
For angled liquid bridges with fixed volume and rod radii, Fig. 4 of the main text illustrates the two possible unfoldings of the pitchfork (in red), with the tilt angle determining which of them is obtained.
Stability changes occur at turning points for imperfect pitchforks. 
They are characterized by $\partial\lambda/\partial a=0$ and the equivalent condition $\partial G/\partial a=0$ when $a$ solves Eq. \ref{eq:bifeq}.
Therefore, the maximum length of angled liquid bridges, $\lambda_\text{max}$, is determined by the equation $\partial G/\partial a=0$. 

To find analytic expressions for the maximum length and threshold angle, we need to know the relative strengths of the parameters in the experiments.
From their typical values in the experiments, we obtain $h\sim\mathcal{O}(\epsilon^{2})$, $\phi\sim\mathcal{O}(\epsilon)$, and $v\sim\mathcal{O}(\epsilon)$, where $\epsilon\sim 10^{-1}$.
In the complete bifurcation equation (Eq. \ref{eq:bifeq}), we replace $h$, $\phi$, and $v$ with $h \epsilon^2$, $\phi \epsilon$, and $v\epsilon$, respectively. 
This allows us to treat $h$, $\phi$, and $v$ as being $\mathcal{O}(1)$, while the new variable, $\epsilon$, will help us in accurately tracking the order of corrections.

The maximum length of the bridge depends on $h$, $\phi$, and $v$. 
We assume the expansion
\begin{equation}
\lambda_{\text{max}}=\lambda_{1}\epsilon+\lambda_{2}\epsilon^{1+q_{2}}+\lambda_3\epsilon^{1+q_2+q_3}+\cdots,
\end{equation}
where $\lambda_1$, $\lambda_2$, $\lambda_3$, $\cdots$ are functions of $h$, $\phi$, and $v$, with overall orders of $\epsilon$, $\epsilon^{1+q_2}$, $\epsilon^{1+q_2+q_3}$, $\cdots$, respectively.
We enforce $\lambda_{1}$, $\lambda_{2}$, $\lambda_{3},\cdots\neq 0$.
With these substitutions in Eq. \ref{eq:bifeq}, we obtain
\begin{multline}
G_{\text{max}}(\epsilon;a)\coloneq G(\lambda_{\text{max}},\epsilon^{2}h,\epsilon\phi,\epsilon v;a)=
a^{3}\frac{3 \pi }{2 \sqrt{2 \pi ^2+2}}
+\epsilon ^2 \left(\frac{2 h}{\sqrt{2 \pi ^2+2}}-\frac{\phi ^2}{4 \sqrt{2 \pi ^2+2}}\right)
\\
+\epsilon ^3 \left(-\frac{8 \pi  h \lambda_{1}}{\left(2 \pi^2+2\right)^{3/2}}
+\frac{\lambda_{1} \phi ^2}{8\pi \sqrt{2}~  \left(1+\pi ^2\right)^{3/2}}
+\frac{5 \pi  \lambda_{1} \phi ^2}{8 \sqrt{2}~ (1+\pi ^2)^{3/2}}
-\frac{v \phi ^2}{8 \sqrt{2 \pi ^2+2}}\right)
+a\epsilon^{1+q_{2}}\left( \frac{3\lambda_{2}}{\sqrt{ 2\pi^{2}+2 }} \right)
\\
+a\epsilon \left( \frac{3 \lambda_{1}}{\sqrt{2 \pi ^2+2}}-\frac{\pi  v}{\sqrt{2 \pi ^2+2}} \right) 
+a\epsilon^{2}\Big( \frac{27 \lambda_{1}^2}{8\pi \sqrt{2} ~ \left(1+\pi ^2\right)^{3/2}}-\frac{21 \pi  \lambda_{1}^2}{8 \sqrt{2}~ \left(1+\pi ^2\right)^{3/2}}+\frac{11 \pi  v^2}{8 \sqrt{2\pi^{2}+2}}\\
-\frac{7 \pi ^2 \lambda_{1} v}{4 \sqrt{2}~ \left(1+\pi ^2\right)^{3/2}}-\frac{15 \lambda_{1} v}{4 \sqrt{2}~ \left(1+\pi ^2\right)^{3/2}}-\frac{\phi ^2}{4 \pi  \sqrt{2 \pi ^2+2}} \Big) +\cdots=0,
\end{multline}
where only coefficients of $a^{i}\epsilon^{j}$ for which $i+j\leq 3$ have been computed.

The stability criterion established above determines $\lambda_1$, $\lambda_2$, $\lambda_3$, $\cdots$ by requiring $\partial G_\text{max}/\partial a$ to vanish at every order. 
This derivative is given by
\begin{multline}
\frac{ \partial G_{\text{max}} }{ \partial a } =a^{2} \frac{9\pi}{2\sqrt{ 2\pi^{2}+2 }}+\epsilon \left(- \frac{\pi v}{\sqrt{ 2\pi^{2}+2 }}+ \frac{3\lambda_{1}}{\sqrt{ 2\pi^{2}+2 }} \right)
+\epsilon^{1+q_{2}} \frac{3\lambda_{2}}{\sqrt{ 2\pi^{2}+2 }} \\
+\epsilon^{2}\Big( \frac{11\pi v^{2}}{8\sqrt{ 2\pi^{2}+2 }}- \frac{15v\lambda_{1}}{4\sqrt{ 2 }~(1+\pi^{2})^{3/2}}- \frac{7\pi^{2}v\lambda_{1}}{4\sqrt{ 2 }~(1+\pi^{2})^{3/2}}\\
+ \frac{27\lambda_{1}^{2}}{8\pi\sqrt{ 2 }~(1+\pi^{2})^{3/2}}- \frac{21\pi\lambda_{1}^{2}}{8\sqrt{ 2 }~(1+\pi^{2})^{3/2}}- \frac{\phi^{2}}{4\pi\sqrt{ 2\pi^{2}+2 }} \Big)+\cdots.
\end{multline}
From $G_{\text{max}}(\epsilon;a)=0$, we find the amplitude $a$.
This amplitude is the $a^*$ of the main text because it is being evaluated at $\lambda_\text{max}$.
We find the root $a(\epsilon)$ as a Puiseux expansion \cite{puis}:
\begin{equation}
a(\epsilon)=c_{1}\epsilon^{\gamma_{1}}+c_{2}\epsilon^{\gamma_{1}+\gamma_{2}}+c_{3}\epsilon^{\gamma_{1}+\gamma_{2}+\gamma_{3}}+\cdots,
\end{equation}
where $c_{1},c_{2},c_{3},\cdots$ are non-zero coefficients.
The coefficients and exponents are obtained using the Newton-Puiseux algorithm detailed in Ref. \cite{puis}.

At $\mathcal{O}(\epsilon)$, we find that $ \partial G_{\text{max}}/ \partial a $ vanishes only when $\lambda_{1}=\pi v/3$, and this is the maximum length at leading order. 
Continuing with the algorithm, we obtain $  q_2=1/3$ and $\lambda_{2}=- \frac{1}{4} \left( \frac{3\pi}{2} \right) ^{1/3}(8h-\phi^{2})^{2/3}$, with $ \partial G_{\text{max}}/ \partial a $ vanishing at $\mathcal{O}(\epsilon^{4/3})$.
Therefore, $\lambda_{\text{max}}$ is
\begin{equation}\label{eq:maxlength}
    \lambda_{\text{max}}= \frac{\pi v}{3}\epsilon- \frac{1}{4} \left( \frac{3\pi}{2} \right) ^{1/3}(8h-\phi^{2})^{2/3}\epsilon^{4/3}+\cdots.
\end{equation}

In addition to the maximum length, the algorithm yields $a(\epsilon)$, and thus the shape (Eq. \ref{eq:lsshape}), to leading order:
\begin{equation}\label{eq:shapehint}
    a=\epsilon^{2/3} \left( \frac{1}{12\pi} \right) ^{1/3}(8h-\phi^{2})^{1/3},\quad
    (1+f)(x,\theta)=1+\epsilon^{2/3}\left( \frac{1}{12\pi} \right) ^{1/3}(8h-\phi^{2})^{1/3} \sin x.
\end{equation}
This equation implies that the tilt angle plays a stronger role in establishing the neck of the bridge than in breaking the axisymmetry of the overall shape. 
It also hints that the amplitude of the sine, $8h-\phi^2$, plays a crucial role in determining the threshold angle.

\subsubsection{Threshold angle for a symmetric bridge}
To the extent achievable, we want to ensure a symmetric shape of the bridge about the mid-plane $x=0$ at the threshold angle. 
Our strategy to find this angle will be to fine-tune $\phi$ to push the contributions from asymmetric shapes in Eq. \ref{eq:shapeseries} to higher orders in $\epsilon$.
We use the same relative strengths for $h$ and $v$, $h\sim\mathcal{O}(\epsilon^{2})$ and $v\sim\mathcal{O}(\epsilon)$. 

Let $\phi_{\text{thresh}}$ be the threshold tilt angle,
\begin{equation}
    \phi_{\text{thresh}}=\phi_{1}\epsilon+\phi_{2}\epsilon^{1+p_{2}}+\cdots,
\end{equation}
We enforce $\phi_{1}$, $\phi_{2}$, $\cdots\neq 0$.
The maximum length (Eq. \ref{eq:maxlength}) is now slightly modified to incorporate the expansion for the threshold angle,
\begin{equation}
    \lambda_{\text{max}}= \frac{\pi v}{3}\epsilon- \frac{1}{4} \left( \frac{3\pi}{2} \right) ^{1/3}(8h-\phi_{1}^{2})^{2/3}\epsilon^{4/3}+\cdots,
\end{equation}
where $\phi$ in Eq. \ref{eq:maxlength} is replaced by $\phi_{1}$ at order $\epsilon^{4/3}$. 
Substituting the relative strengths of the parameters and the maximum length into the expansion for the shape (Eq. \ref{eq:shapeseries}), we obtain
\begin{equation}\label{eq:eshape}
    (1+f)(x,\theta;\epsilon)=1+a \sin x +a^{2}\left( - \frac{1}{4}+ \frac{1}{4}\cos 2x \right) 
+ \epsilon\left( \frac{v}{3}+\frac{v}{3}\cos x \right) +\cdots,
\end{equation}
where $a=a(\epsilon)$ is a solution of the bifurcation equation,
\begin{multline}
G_{\text{thresh}}(\epsilon;a)\coloneq G(\lambda_{\text{max}},\epsilon^{2}h,\phi_{\text{thresh}},\epsilon v;a)=
a^{3} \frac{3\pi}{2\sqrt{ 2\pi^{2}+2 }}
+\epsilon^{2}\left( \frac{2 h}{\sqrt{2 \pi ^2+2}}-\frac{\phi_{1}^{2}}{4 \sqrt{2 \pi ^2+2}} \right)\\
-\epsilon^{2+p_{2}}\frac{\phi_{1}\phi_{2}}{2 \sqrt{2 \pi ^2+2}}
-\epsilon^{2+2p_{2}}\frac{\phi_{2}^{2}}{4 \sqrt{2 \pi ^2+2}}
+\epsilon^{3}\left( -\frac{8 \pi ^2 h v}{3 \left(2 \pi ^2+2\right)^{3/2}}+\frac{(5 \pi ^2+1) v \phi_{1}^{2}}{24 \sqrt{2} \left(1+\pi ^2\right)^{3/2}}-\frac{v \phi_{1}^{2}}{8 \sqrt{2 \pi ^2+2}} \right) \\
-a\epsilon^{4/3} \left( \frac{3\pi}{2} \right) ^{1/3}\left( \frac{3(8h-\phi_{1}^{2})^{2/3}}{4\sqrt{ 2\pi^{2}+2 }} \right)
+a\epsilon^{2}\left( \frac{\pi v^{2}}{2\sqrt{ 2\pi^{2}+2 }}- \frac{\phi_{1}^{2}}{4\pi \sqrt{ 2\pi^{2}+2 }} \right)
+\cdots =0.
\end{multline}
We repeat the Newton-Puiseux algorithm to find $a(\epsilon)$.
We assume an expansion for $a(\epsilon)$,
\begin{equation}
a(\epsilon)=c_{1}\epsilon^{\gamma_{1}}+c_{2}\epsilon^{\gamma_{1}+\gamma_{2}}+c_{3}\epsilon^{\gamma_{1}+\gamma_{2}+\gamma_{3}}+\cdots,
\end{equation}
where $c_{1}$, $c_{2}$, $\cdots\neq 0$. 
We find that $\phi_{1}=\sqrt{ 8h }$ pushes the asymmetry of the bridge beyond $\mathcal{O}(\epsilon^{2/3})$, and this is our threshold angle to leading order.
This confirms our remark about the shape of the bridge in Eq. \ref{eq:shapehint}. 

The projection hypothesis, Eq. 1 of the main text, predicts the threshold angle to occur when the projected areas of the two rods become equal:
\begin{equation}
    \cos \phi_\text{proj}-\frac{R_1^2}{R_2^2} =0,
\end{equation}
where $R_1$ and $R_2$ are the radii of the level and tilted rods, respectively.
Rewriting this condition for $\phi_\text{thresh}$ and using $h$, we find
\begin{align}
    \cos(\phi_{1}\epsilon+\phi_{2}\epsilon^{1+p_2}+\dots)-\left(\frac{1-\epsilon^2 h}{1+\epsilon^2 h}\right)^2 &=0 \\
   \implies  \left(4h -\frac{\phi_1^2}{2}\right)\epsilon^2 +\mathcal{O}(\epsilon^{2+p_2})&=0.
\end{align}
Therefore, to leading order, the threshold angle predicted by the projection hypothesis is $\phi_1=\sqrt{8h}$, which is equivalent to the first order threshold angle obtained above.

Proceeding with the algorithm, we find that it is possible to push the contributions from asymmetric shapes beyond $\mathcal{O}(\epsilon)$ when $p_{2}=1$ and $\phi_{2}=- \frac{\sqrt{ 2h }}{3}v$.
Thus, we have obtained the threshold angle at which the bridge possesses symmetry about the mid-plane $x=0$ up to $\mathcal{O}(\epsilon)$,
\begin{equation}\label{eq:thresh}
    \phi_{\text{thresh}}=\epsilon\sqrt{ 8h }-\epsilon^{2} \frac{\sqrt{ 2h }}{3}v+\dots.
\end{equation}
On either side of this threshold angle, the shape of an angled bridge at the length at which it breaks is asymmetric about the mid-plane.
This asymmetry is carried over to the post-rupture volume distributions.

\end{document}